\documentclass[letterpaper,12pt]{article}
\usepackage{amsmath}
\usepackage{graphicx,psfrag,epsf}
\usepackage{textcomp}
\usepackage{color}

\usepackage[utf8]{inputenc} 
\usepackage[T1]{fontenc}    
\usepackage{hyperref}       
\usepackage{url}            
\usepackage{booktabs}       
\usepackage{amsfonts}       
\usepackage{nicefrac}       
\usepackage{microtype}      
\usepackage{lipsum}

\usepackage{graphicx}
\usepackage{caption}
\usepackage{subcaption}
\usepackage[shortlabels]{enumitem}

\usepackage{natbib} 

\title{TRAP: A Predictive Framework for Trail Running Assessment of Performance}

\author{
  Riccardo Fogliato, Natalia L. Oliveira, Ronald Yurko\\
  Department of Statistics \& Data Science\\
  Carnegie Mellon University\\
  Pittsburgh, PA 15213 \\
  \texttt{\{rfogliat, nlombard, ryurko\}@andrew.cmu.edu}
}

\begin{document}
\maketitle


\begin{abstract}
Trail running is an endurance sport in which athletes face severe physical challenges. Due to the growing number of participants, the organization of limited staff, equipment, and medical support in these races now plays a key role. Monitoring runner's performance is a difficult task that requires knowledge of the terrain and of the runner’s ability. In the past, choices were solely based on the organizers’ experience without reliance on data. However, this approach is neither scalable nor transferable. Instead, we propose a firm statistical methodology to perform this task, both before and during the race. Our proposed framework, Trail Running Assessment of Performance (TRAP), studies (1) the the assessment of the runner’s ability to reach the next checkpoint, (2) the prediction of the runner’s expected passage time at the next checkpoint, and (3) corresponding prediction intervals for the passage time. We apply our methodology, using the race history of runners from the International Trail Running Association (ITRA) along with checkpoint and terrain-level information, to the “holy grail” of ultra-trail running, the Ultra-Trail du Mont-Blanc (UTMB) race, demonstrating the predictive power of our methodology.
\end{abstract}

\textit{Keywords:} Trail running, UTMB, ITRA, random forests.

\section{Introduction}

What is trail running? According to the \citet{camb}, trail running is the sport or activity of running along trails. More generally, the term may also include the activity of running on trails possibly with sections of paved road, although this is sometimes referred to as mountain running. The term is often used interchangeably with cross country (xc) running, although the latter typically refers specifically to competitive running. Last, trail running differs from ultra running, that is running for distances longer than a marathon (42.195 km); indeed, ultra-trail running is a special case of both ultra running and trail running.
However, these definitions turn out to be rather technical and specific when compared to the broader viewpoints of race organizers, sponsors, and media outlets. According to the International Trail Running Association (ITRA, \url{https://itra.run/}), the largest trail running organization worldwide, trail running is a sport that takes place in a natural environment and enhances a shared sense of community among runners. Other definitions of trail running emphasize aspects such as the runner's connection with nature, the respect for each other, the challenges, risks, and freedom.
Trail running, which in 2015 was recognized by the International Association of Athletics Federations (IAAF) as a track and field sport~\citep{iaaf15}, embodies all these definitions.\\

\begin{figure}[t]
\begin{subfigure}{.5\textwidth}
  \centering
  \includegraphics[width=\textwidth]{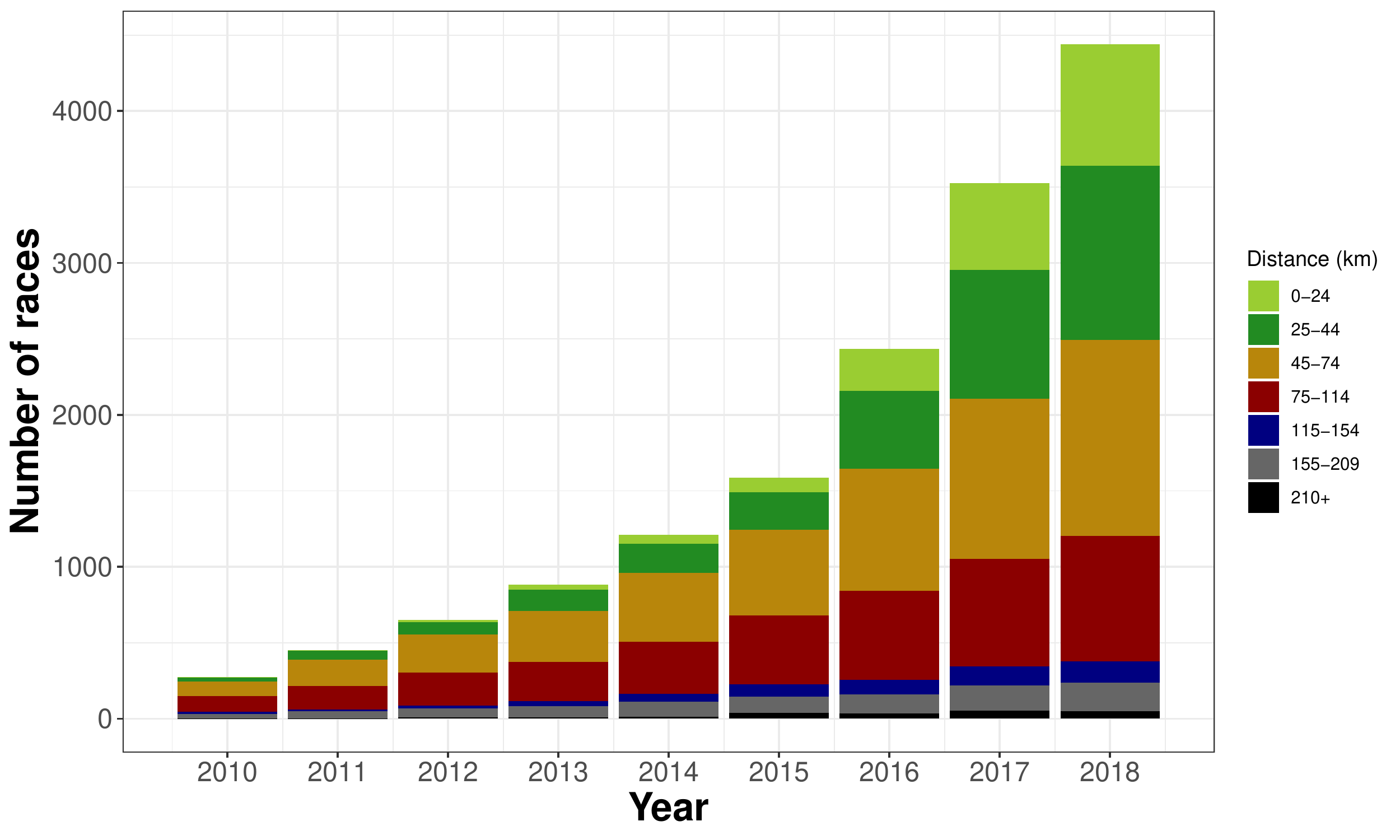}
  \caption{}
  \label{fig:nraces_ITRA}
\end{subfigure}%
\begin{subfigure}{.5\textwidth}
  \centering
  \includegraphics[width=\textwidth]{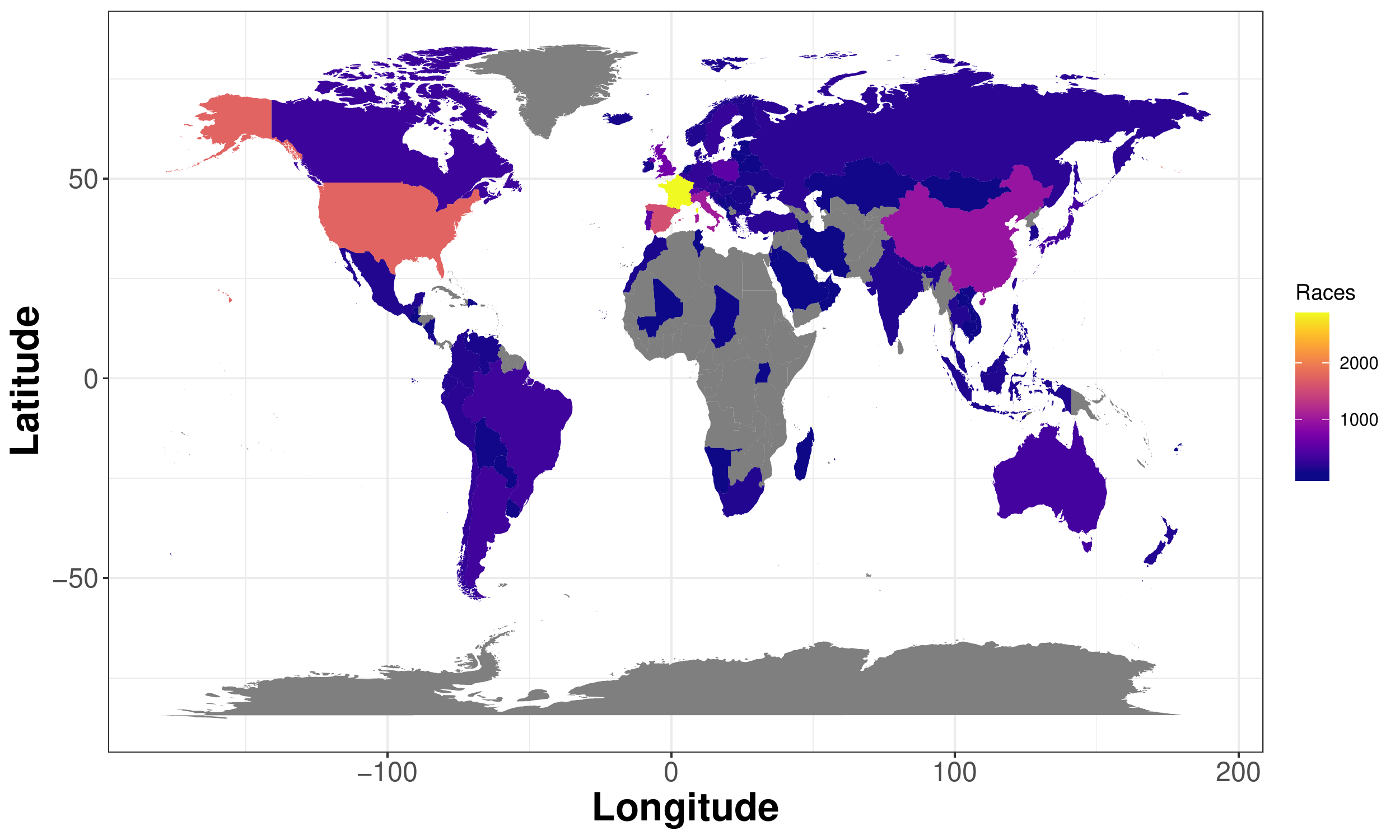}
  \caption{}
  \label{fig:country_races_ITRA}
\end{subfigure}
\caption{Number of races listed on the International Trail Running Association's (ITRA) website from 2010 until 2018 separated by (a) year and distance category; (b) country. Data is taken from the ITRA data set. }
\end{figure}

Although all these terms may sound familiar today, they would have definitely sounded less familiar ten years ago. Indeed, the world of trail running has experienced an exponential growth in terms of the number of athletes, races, news and media coverage~\citep{ives15, burgess15, saiidi15}, and interest in the sport industry only in the last decade~\citep{anger15}. Despite the recent surge in popularity, trail running has been practiced for decades. One of the most famous trail running competitions, the Western States Endurance Run (\url{https://www.wser.org/}), a 161 km race with 5500 m\footnote{Throughout the paper, we will use ``m'' and ``km'' to indicate meters and kilometers respectively.} of elevation gain in California, was run for the first time in 1974; at that time, it was only known as a horse race and a participant decided to run instead of riding his horse. In 2019, race organizers used a lottery system to select approximately 300 participants out of a pool of over 5000 applicants. Similarly, Ultra-Trail du Mont-Blanc (UTMB, \url{https://utmbmontblanc.com/}), with 171 km and 10000 m of elevation gain crossing France, Italy, and Switzerland, started with only a few participants in 2003 and now manages admissions through a lottery system. The cases of Western States and Ultra-Trail du Mont-Blanc are not isolated: obtaining admission to the most well known races is becoming increasingly harder over time. However, the number and the spectrum of characteristics of trail running races is also widening. Figure~\ref{fig:nraces_ITRA} shows the number of races registered in the ITRA circuit from 2010 to 2018. The number of races recorded on the website grew sixteen times in only eight years, from 275 races in 2010 to 4439 in 2018. Although this estimate is affected by selection bias, as the participation to the ITRA circuit has also been growing over time, it still provides some insight into the rise in popularity of trail running. The stronger presence of ITRA in the European continent is noticeable in Figure~\ref{fig:country_races_ITRA}: the country with the highest number of races in the 2010-2018 period is France, followed by the United States, Spain, Italy, and China.\\

The current popularity of trail running is likely due to a variety of factors, among which the extreme nature of this sport. Indeed, most of the famous races, such as UTMB, Western States, Hardrock 100 (\url{https://hardrock100.com/}), Lavaredo Ultra-Trail (LUT, \url{https://www.ultratrail.it}), Tor des G\'eants (TOR, \url{https://www.tordesgeants.it}), Marathon des Sables (\url{https://www.marathondessables.com/en}), and Ultra-Trail Mount Fuji (\url{https://www.ultratrailmtfuji.com/en/}) are all ultra-trails. Qualification for these races requires runners to have already completed races of comparable difficulty. For instance, in 2019, runners needed ten ITRA points\footnote{The ITRA points are the scoring system adopted by ITRA. We describe ITRA and the scoring system in our supplementary materials.} in at most two races only to enter UTMB's lottery. Still, even among these highly skilled runners, the fraction of dropouts in the races is large. Table~\ref{tab:races_info} shows that, for some of the races mentioned above, this percentage ranges from 20\% to 40\%, which means that on average one or two out of five runners do not conclude the race. Why such a large fraction of runners drops out? Intuitively, dropouts are proportional to the difficulty of the race in terms of finishing time, elevation gain, distance, and environment. For example, longer time spent in the mountains is linked to a higher likelihood of incidents: runners are more likely to suffer from dehydration, hypothermia, injuries, and sleep deprivation due to fatigue. We explore these relationships in Section~\ref{sec:trap}. Although these famous races get more visibility in the media, most of the races feature shorter distances, as shown in Figure~\ref{fig:nraces_ITRA}. 
In these cases, some of the risks (e.g., sleep deprivation) are mitigated, but it is still important to fully evaluate the threats imposed by the environment (e.g., weather) in order to minimize the likelihood of tragic events and manage the organization of resources along the trail.\\

\begin{table}[!b]
\begin{centering}
\caption{Distance (D), elevation gain (D+), participants, finishers, and maximum time allowed relative to the 2018 editions of Lavaredo Ultra-Trail (LUT), Western States Endurance Run (Western States), Ultra-Trail du Mont-Blanc (UTMB), Hardrock 100 (Hardrock), Tor des G\'eants (TOR), Ultra-Trail Mt. Fuji.}
\label{tab:races_info}
\begin{tabular}{ |c|c|c|c|c|c|c| } 
\hline
Race & D (km) & D+ (m) & Runners & \% Finishers & Time barrier (h) \\
\hline
LUT & 120 & 5760 & 1608 & 74\% & 30 \\
Western States & 161 & 5360 & 369 & 81\% & 30 \\
UTMB & 171 & 9930 & $\sim$2300 & 69\% & 45\\
Hardrock 100 & 160 & 10365 & 146 & 61\% & 48\\
TOR & 330 & 24000 & $\sim$750 & $\sim$70\% & 150 \\
Ultra-Trail Mt. Fuji & 168 & 9500 & 1480 & 73\% & 46 \\
\hline
\end{tabular}
\end{centering}
\end{table}

Organizing trail running races requires extensive knowledge of the terrain and of the runners' skills. For instance, organizers need to detect runners that are experiencing difficulties and consequently either contact them or take action. Moreover, they need to properly allocate medical staff, equipment, volunteers, buses, food, and drinks at aid stations. The current organizational approach is problematic for the following three reasons. First, knowledge of the terrain and of the runner's skills is either crowdsourced from several individuals with different areas of expertise or even obtained from just a handful of organizers. Second, this knowledge is transferred across races via heuristics. Third, runners' ability is typically estimated, again via heuristics, solely based on past (if available) and current performance in that race and only during the race. Such an approach is neither scalable nor transferable: knowledge does not improve with data and cannot be transferred across races and individuals.\\

While we understand that the setting of trail running may sometime require knowledge that goes beyond the information that can be recorded in the data, we recognize the importance of introducing a data-driven benchmark that supports the organization of these races. To our knowledge, such a benchmark does not currently exist. The main contributions of this paper are the following:
\begin{itemize}
    \item The collection of data of over 1.7 million runners in more than fifteen thousand races from the International Trail Running Association (ITRA). This rich data set is, to our knowledge, the largest data set on trail running ever collected for research purposes. All collected data used in this manuscript are available at a GitHub repository \texttt{https://github.com/ricfog/TRAP\_data}. A detailed description and exploratory analysis of the data is provided in the supplementary materials. 
    \item A novel framework, Trail Running Assessment of Performance (\textbf{TRAP}), for assessing runners' performance both before and during the race. Leveraging statistical machine learning tools, our methodology targets three quantities related to the runner's
\begin{enumerate}[(a)]
    \item expected passage time; 
    \item probability of dropping out;
    \item prediction interval for passage time.
\end{enumerate}
These quantities are computed for the following checkpoint (i.e., the locations where runners have their bibs scanned) based on the runner's position. However, the framework is readily extensible to multiple checkpoints ahead at arbitrary positions. We present the TRAP framework through a case study of the 2015, 2016, 2017, and 2018 editions of the Ultra-Trail du Mont-Blanc (UTMB) race. 
\end{itemize}

The paper is structured as follows. Section~\ref{sec:related} describes related work. Section~\ref{sec:utmb} provides an overview of the joined data from ITRA and our case study UTMB. Section~\ref{sec:trap} contains description and application of the TRAP methodology on UTMB with the various model results presented in Section~\ref{sec:results}. Section~\ref{sec:discussion} contains final discussion and directions for future work. \\



\section{Related work}\label{sec:related}
The literature on trail running inherits results from two large bodies of works on ultra-trail running and track (or road) running. However, most of these studies focus on the medical aspects of these activities and therefore their goals clearly differ from the purpose of our paper. Although this literature would deserve an extensive review, we limit the discussion in this section to the papers whose settings and scopes are closest to ours.\\

Researchers have thoroughly investigated the role of pacing in running, that is how athletes distribute the consumption of their own energy throughout the race. 
We start by reviewing results on marathons. Multiple studies report that men are more likely to slow their pace compared to women, even with several different definitions of pacing. \citet{march2011age} study the effect of age, sex, and finish time on marathon pacing, which they define as the ratio of the mean speed in the last 9.7 km and in the first 32.5 km. They find that older, faster, and female runners are more likely to keep the pace steady compared to younger, slower, and male runners respectively. Using a similar definition of pacing, \citet{trubee2014effects} analyse the 2007 and 2009 editions of the Chicago marathon and show that age, sex, finish times, and heat stress are significant predictors of pacing; in particular, ceteris paribus, heat stress would lead to greater variability in pacing. \citet{haney2011description} use data of the Las Vegas and San Diego marathons uploaded on Garmin Connect and show that pace, defined as the ratio of the standard deviation and the mean of the total duration, has less variability for faster rather than for slow runners.
\citet{santos2014influence} analyse a large data set containing the results of several editions of the New York City marathon and they find that both men and women try to avoid fast starts in order to keep their pace constant in the second part of the race. 
Studying fourteen marathons in the US during 2011, \citet{deaner2015men} find that, after controlling for a variety of factors, men seem to be more likely than women to decrease their pace, defined as the ratio of finish times of the second and first half marathons.
In an analysis of a data set of one million participants in several marathons, \citet{krawczyk2017we} find that men, youngest, and oldest participants tend to slow down more in the second half of the marathon. This finding partially matches those of~\citet{march2011age} and~\citet{trubee2014effects} mentioned above. Similar findings have been reported by~\citet{cuk2020sex, nikolaidis2017effect, nikolaidis2019performance}.
In a large-scale analysis of over 1.7 million recreational runners, \citet{smyth2018fast} find that women tend to pace the race more evenly than men. 
Pacing also seems to be tied to optimism. 
\citet{krawczyk2017we} analyse predicted and true finish times for the 2012 Warsaw marathon, and they find that optimism is associated with more slowdown in the second half of the marathon.
\citet{hubble2016gender} use data from the 2013 Houston Marathon to show that men are more overconfident in their future performance compared to women and their pace decreases more substantially in later stages of the marathon. \\

The body of work on long distance races is less rich, but nonetheless contains partial answers to some of the questions raised in the marathon setting. For example, using data from the 1995 IAU World-Championship, a 100 km race organized by the International Association of Ultrarunners (IAU), \citet{lambert2004changes} find that all participants decreased their speed during the race, with faster runners showing a smaller and delayed decrease. 
However, empirical evidence of an effect of pacing on performance is mixed, with \citet{hoffman2014pacing} and \citet{kerherve2016pacing} reporting positive and null results respectively.
\citet{bartolucci2015finite} fit a finite mixture latent trajectory model on runners' performance in the 2013 IAU World-Championship, a 24-hour continuous race. Interestingly, their results show that runners are more likely drop out in the middle of the race; we observe a similar pattern in our case study on UTMB. They also observe that runners' speed gradually decreases only to increase towards the end of the race, similarly to what was reported by \citet{lambert2004changes}. Pacing has also been shown to depend on age and experience \citep{rust2015non, knechtle2015pacing}.\\

The roles of gender and age as predictors of performance have also been analysed. \citet{peter2014sex} use 35 years of several 24-hour ultramarathons held around the world to study gender gaps in performance. The authors conclude that, although the time gap between the top men and women has been decreasing over the years, it is unlikely that the gap will be closed in the near future. In addition, they find that ultramarathoners typically achieve their best performance after the age of 35. \citet{zingg2014will} reach a similar conclusions on an analysis of 50 years of ultramarathons from 50 km to 1000 km long. \citet{coast2004gender} analyse world records for races from 100 m to 200 km long and they conclude that the gender gap is larger for longer events; however, they warn that this result may be confounded by the fact that only a few women participate in long distance races.\\

A different line of work has focused on the relationship between performance and demographics or contextual factors; most of these studies belong to the medical literature. Here we report only a few of them. 
\citet{ely2007impact} study the impact of weather on performance. Using longitudinal marathon and weather data for seven cities, they find that an increase in temperature from five to twenty-five degrees Celsius leads to lower levels of performance in the marathon. 
Similarly, \citet{knechtle2014prediction} use linear regression to model finish times in half marathons as a function of body fat measure and running speed during training; they test the coefficients of the corresponding equation proposed in the literature \citep{knechtle2011predictor, rust2011predictor, friedrich2014comparison}. In a review of the literature, \citet{keogh2019prediction} identified 114 of these equations in thirty-six studies. \\

Our study is close to the body of work on prediction of finish times, but focuses on arbitrary distances. In addition, our measures of interest have not been discussed yet in the literature. Our work is, to our knowledge, novel both in terms of the setting (trail running) and the methodology (TRAP). However, we are aware of a service provided by a proprietary software, LiveTrail (\url{https://www.livetrail.net/}), similar to part of the framework presented here. LiveTrail offers the technology and equipment for race organizers and supporters to monitor runners' performance during the race. Elite runners are typically given GPS devices that allow for real-time tracking of their positions, while all other runners get their bibs scanned only at checkpoints. After each scanning, LiveTrail outputs a prediction of the expected passage time for the next aid station. In Section~\ref{sec:trap} we present our open-source algorithms for this task. We believe that this part of the technology is easy to implement through our framework and should be available to race organizers at no costs. We also find that expected passage time is often not informative in the context of trail running, in which prediction tasks are difficult. Moreover, besides the estimate being unreliable, expected time is likely not a measure of central interest because:
\begin{itemize}
    \item supporters (e.g., family, friends, or sponsors) need to be present at the following aid station to help the runner. Therefore they are interested in a \textit{minimum passage time};
    \item race organizers and physicians need to be alerted if a runner has suffered from a serious injury and cannot reach the next checkpoint. Therefore they are interested in a \textit{maximum passage time}.
\end{itemize} 
For this reason, we extend the framework by proposing upper and lower bounds for passage times via prediction intervals, discussed in Section~\ref{sec:trap}.\\

\section{Overview of Ultra-Trail du Mont-Blanc data}\label{sec:utmb}

\begin{figure}[t]
  \centering
  \includegraphics[width=\textwidth]{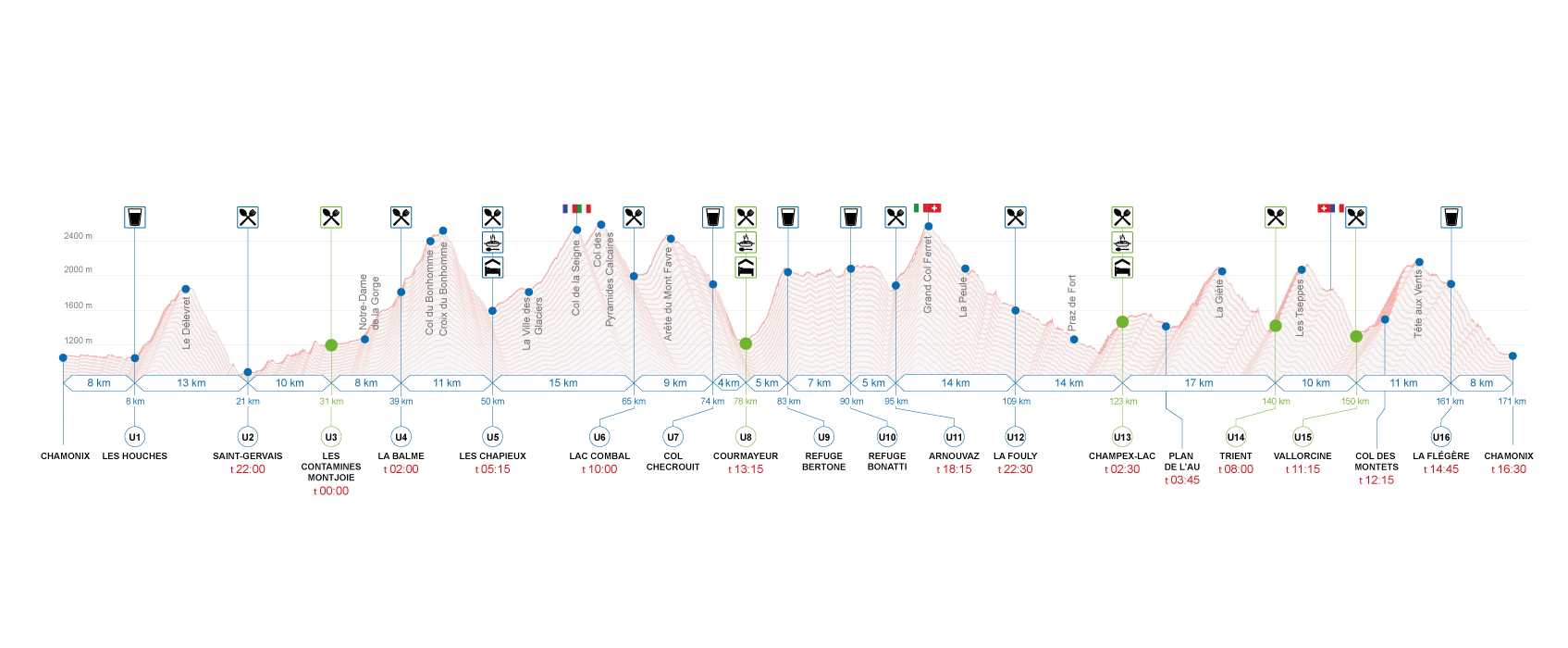}
\caption{Profile of the Ultra-Trail du Mont-Blanc (UTMB) race for the 2018 edition. The profile shows checkpoints (below, in capital letters) with corresponding time barriers (below, in red) and services available (above, in boxes), distances between aid stations (below, in blue boxes) and cumulative (below, in blue). The vertical axis represents the elevation.}
 \label{fig:utmb_profile}
\end{figure}

Ultra-Trail du Mont-Blanc (UTMB) is an ultramarathon\footnote{UTMB refers to both the race and the circuit in which the homonymous race is included. This circuit consists of the UTMB race and also of five other races (YCC, MCC, OCC, CCC, TDS, PTL); these competitions are held in the same week as UTMB itself, they all have the same finish line in Chamonix featuring different distances and elevation gains. We focus on the UTMB race because it's the oldest and most famous among these competitions.} held in France every year during the last week of August. UTMB is part of the Ultra-Trail World Tour, an international circuit of ultra-trail running races. In this race, participants run along the Tour du Mont Blanc hiking path: starting from Chamonix, they cross the borders of Italy and Switzerland, and run back to Chamonix making a complete tour of the Mont Blanc.
Since its first edition in 2003, the race has been modified over the years and the last seven editions (2013-2019) featured a distance of approximately 170 km with over 10000 m of elevation gain; the profile for the 2018 edition is shown in Figure~\ref{fig:utmb_profile}. While elite runners typically complete the race in about twenty hours, most runners cross the finish line in more than forty hours, right before the time barrier of 46.5 hours. The participation is limited to approximately 2500 runners and it requires a minimum of ten ITRA points scored in at most two races. Admission is based on a lottery system. Thanks to its extreme nature and the spectacular views, UTMB is seen by many runners as the pinnacle of a career in trail running. We present the race through an exploratory data analysis in order to lay the groundwork for the TRAP predictive framework in Section~\ref{sec:trap}.\\

The following analysis is based on three separate data sets: (1) the ITRA data set described in our supplementary materials; (2) a second data set, downloaded from Kaggle, containing the results of UTMB for the editions 2015-2017; (3) a third data set with the results of UTMB for the 2018 edition, that we scraped from the race's website. Sources (2) and (3) contain information on passage times at checkpoints throughout the race. Instead, we use source (1) to leverage information on the runner's past race history. Indeed, since UTMB requires ITRA points, all runners competing in the race need to have run at least two races in the ITRA circuit; this ensures that we have prior records for all of them.\footnote{We cannot rule out the possibility that two (or more) runners have the same name and therefore we mistakenly aggregate their past records from the ITRA data set.}  The following analysis is based on the editions of UTMB 2015, 2016, 2017, and 2018. We consider this time window because the path of the course was slightly modified in 2017 and this makes the design of a predictive framework both more interesting and challenging. The total number of participants in our UTMB data set is 8689, with approximately 2600 runners per edition. For each edition, we dropped almost 400 participants because we could not automatically match their names with records from ITRA. The analyses in both this section and in Section~\ref{sec:trap} are based on this reduced data set. \\

Who are UTMB participants? 
The large majority (89-92\%) of runners is male. 
Most runners belong to the age categories V1 (45\% M, 42\% F),\footnote{``M'' and ``F'' inside the parentheses indicate male and female runners respectively. The percentages correspond to the fraction of runners in each category.} SE (32\% M, 38\% F), and V2 (20\% M, 17\% F), which correspond to the age ranges 40-49, 23-40, and 50-59 respectively. The minimum age to compete in the race is 20, with only ten male runners under 23 across all editions.
Interestingly, there are also eleven runners (ten M, one F) that are more than 70 years old.
The most frequent nationalities of the runners are France (34\% of all runners), Italy (9\%), Spain (8\%), UK (7\%), Japan (6\%), United States (4\%), China (3\%), and Poland (3\%). 
Prior to UTMB, runners have completed on average twelve races of the ITRA circuit with mean race distance and elevation of 90 km and 4224 m respectively. The average total distance for all races is 1063 km with 51205 m of elevation. Only 30\% of the runners have already run races with distance equal to or longer than the UTMB's.\\

\begin{figure}[t]
\begin{subfigure}{.5\textwidth}
  \centering
  \includegraphics[width=\textwidth]{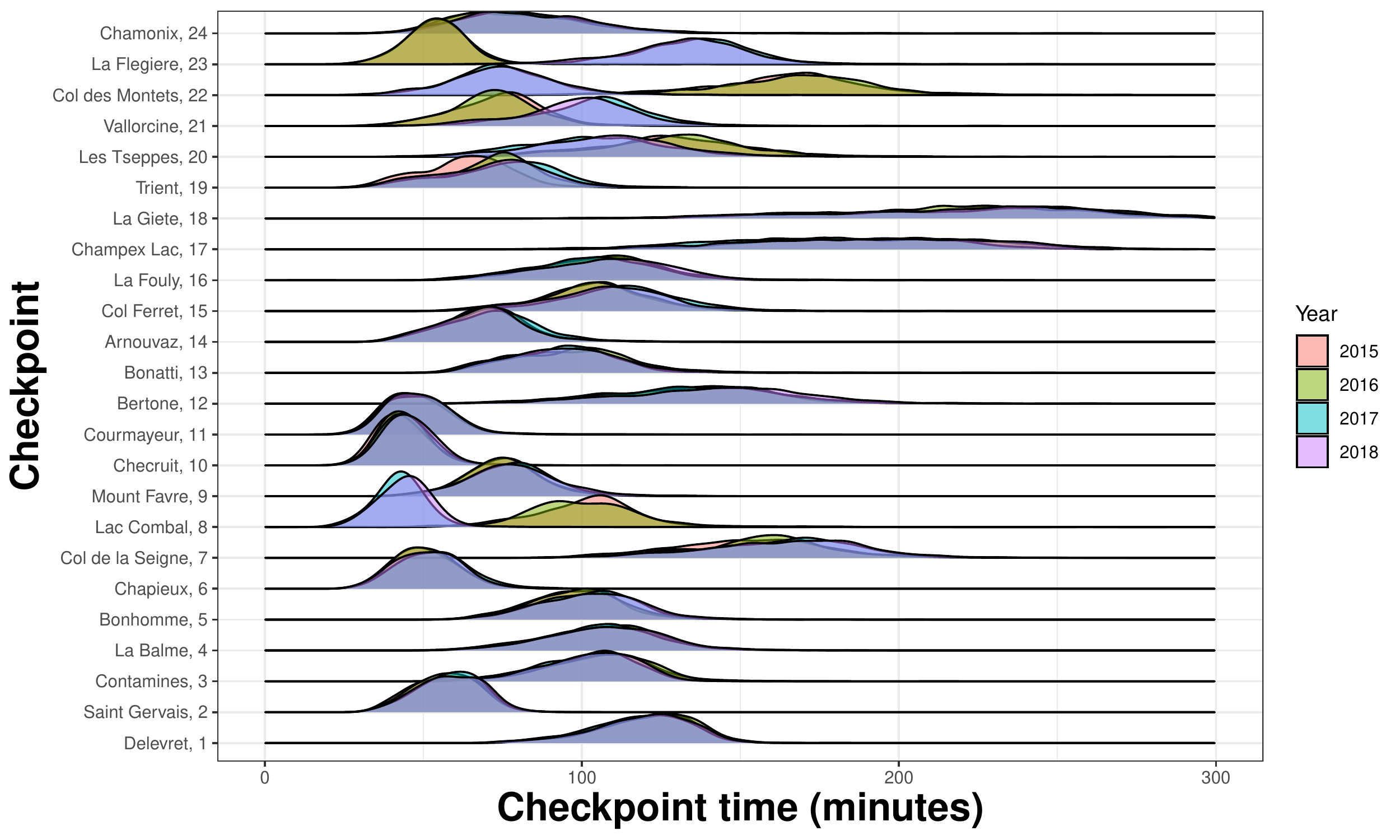}
  \caption{}
  \label{fig:utmb_checkpoint_time}
\end{subfigure}%
\begin{subfigure}{.5\textwidth}
  \centering
  \includegraphics[width=\textwidth]{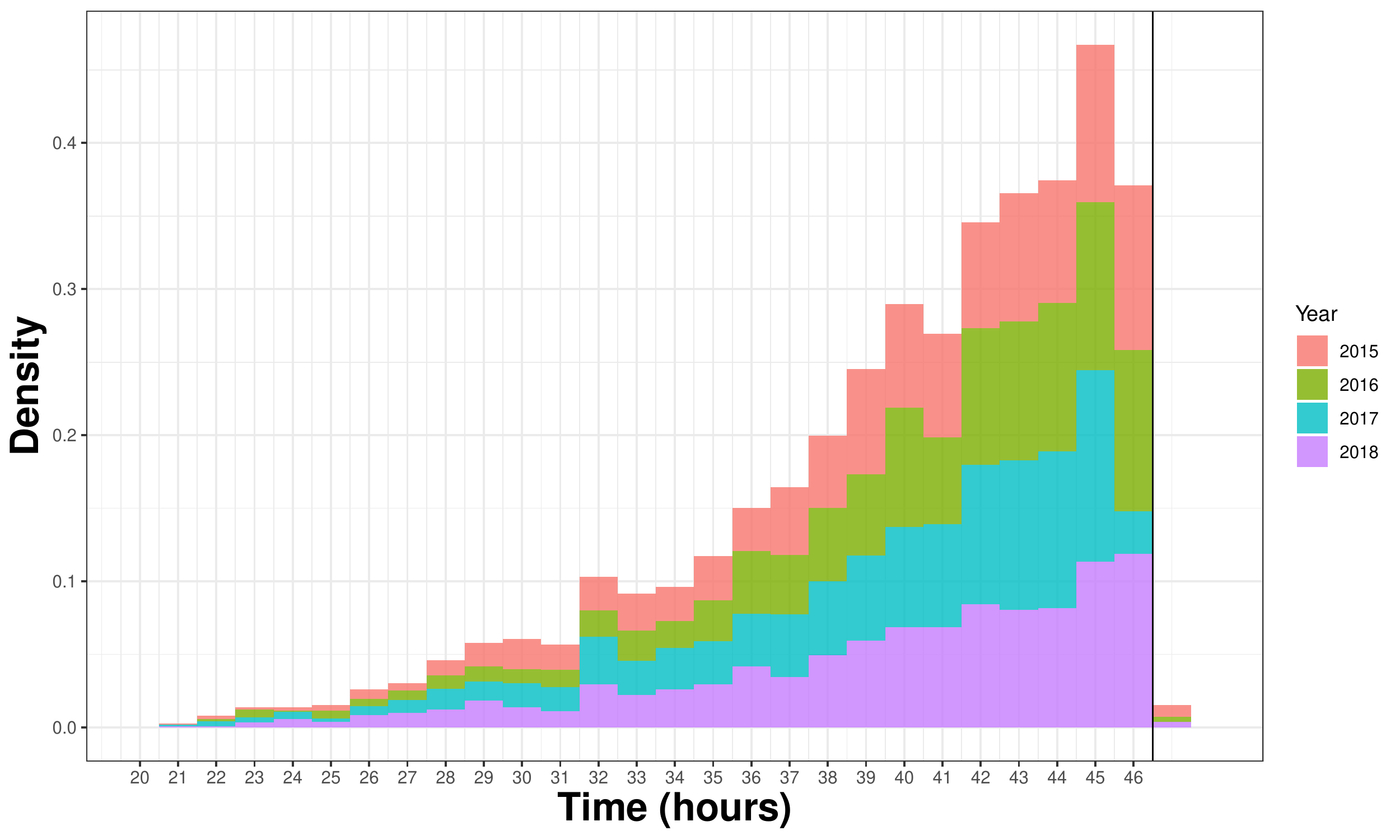}
  \caption{}
  \label{fig:utmb_final_time}
\end{subfigure}
\caption{Data from the editions 2015-2018 of the UTMB race. (a) Density estimates of the distribution of passage times for runners at 24 checkpoints by year. The checkpoints are relative to the 2017-2018 editions. (b) Density histogram for the distribution of arrival times in Chamonix by year.}
\end{figure}

How long does it take for participants to complete the race? The race starts at 18:00 on Friday and the time barrier is at 16:30 on Sunday, which translates into a maximum allowed time of $46.5$ hours to complete the race. Figure~\ref{fig:utmb_final_time} shows the distribution of finish times for the years 2015-2018 through stacked density histograms. While the winner typically crosses the finish line in Chamonix in twenty hours, only 5-7\% and 39-43\% of the runners complete the race in less than thirty and forty hours respectively. The mean arrival time is approximately forty hours and it does not differ substantially across age categories and genders. This means that most athletes complete the race close to the time barrier and consequently spend two nights out on the trails. Besides the final time barrier in Chamonix, there are multiple others located at some of the checkpoints, highlighted in red in Figure~\ref{fig:utmb_profile}. If a runner does not reach the checkpoint before the time barrier, they are forced to withdraw from the race. Time barriers allow course organizers to concentrate resources and better monitor runners along the path. \\

How often is runners' performance recorded? As mentioned above, the path of the race was modified in 2017 and therefore the 2015 and 2016 editions differ from those of 2017 and 2018 in some of the checkpoints positions. For this reason, we restrict the checkpoints considered from all the checkpoints shown in Figure~\ref{fig:utmb_profile} down to only the twenty-four listed on the vertical axis of Figure~\ref{fig:utmb_checkpoint_time}. While such a choice makes the prediction tasks of Section~\ref{sec:trap} harder, it simplifies the presentation of the exploratory data analysis in the current section. Figure~\ref{fig:utmb_checkpoint_time} shows the density estimates for the runners' passage times at the twenty-four checkpoints. Each passage time is calculated as the difference between cumulative passage times (i.e. since the beginning of the race) at two consecutive checkpoints. For example, it takes approximately two hours for most runners to arrive to checkpoint 1 (Delevret) from the start of the race in Chamonix, and from three to four hours to reach 18 (La Giete) from 17 (Champex Lac). These estimates are computed considering only those that have not dropped out yet. There are at least three notable patterns in Figure~\ref{fig:utmb_checkpoint_time}. First, since the four editions of the race differ in the 8th (Lac Combal), 20th (Les Tseppes), and 22nd (Col des Montets) checkpoints, the corresponding partial times do not match. Second, despite different environmental conditions (e.g. weather) and dropout rates (37\%, 42\%, 33\%, 30\% for years 2015-2018 respectively) across the editions, the distributions of runners' passage times remain similar for the checkpoints whose location did not change. Third, mean and skewness of the distributions largely vary across checkpoints. For instance, most runners run the 8 km distance between checkpoints 3 (Les Contamines) and 4 in less than two hours, but it may take them from one to five hours to cover the 14 km distance separating checkpoints 16 (La Fouly) and 17 (Champex Lac). As we show later, this variability is a function of the distance between checkpoints, cumulative distance from the start, and elevation gain.\\

Can runners access food and drinks? UTMB is run in semi-autonomy, which means that runners are required to bring a certain amount of mandatory gear (e.g., jacket, cellphone, food, water), food, and drinks with them, but they can also get extra resources provided by race organization along the course path. While drinks are present at all checkpoints, food is available only at checkpoints 2, 3, 4, 6, 8, 11, 14, 16, 17, 19, 21, 24 as shown in Figure~\ref{fig:utmb_checkpoint_time}. We call these checkpoints ``aid stations''. Beds are available at checkpoints 6, 11, 17, 24. Runners can meet supporters (e.g., family members, friends, or sponsors) and receive additional gear at checkpoints 3, 11, 17, 19, and 21. Runners can also send personal gear to checkpoint 11 prior to the start to the race. Buses provided by the organization take runners that have withdrawn from the race back to Chamonix, and they are available at checkpoints 2, 3, 6, 8, 11, 14, 16, 17, 19, and 21. Medical personnel is present at most of the aid stations and, in case of emergencies, runners can always request help from medical personnel by calling the organization.\\

\begin{figure}[t]
\begin{subfigure}{.5\textwidth}
  \centering
  \includegraphics[width=\textwidth]{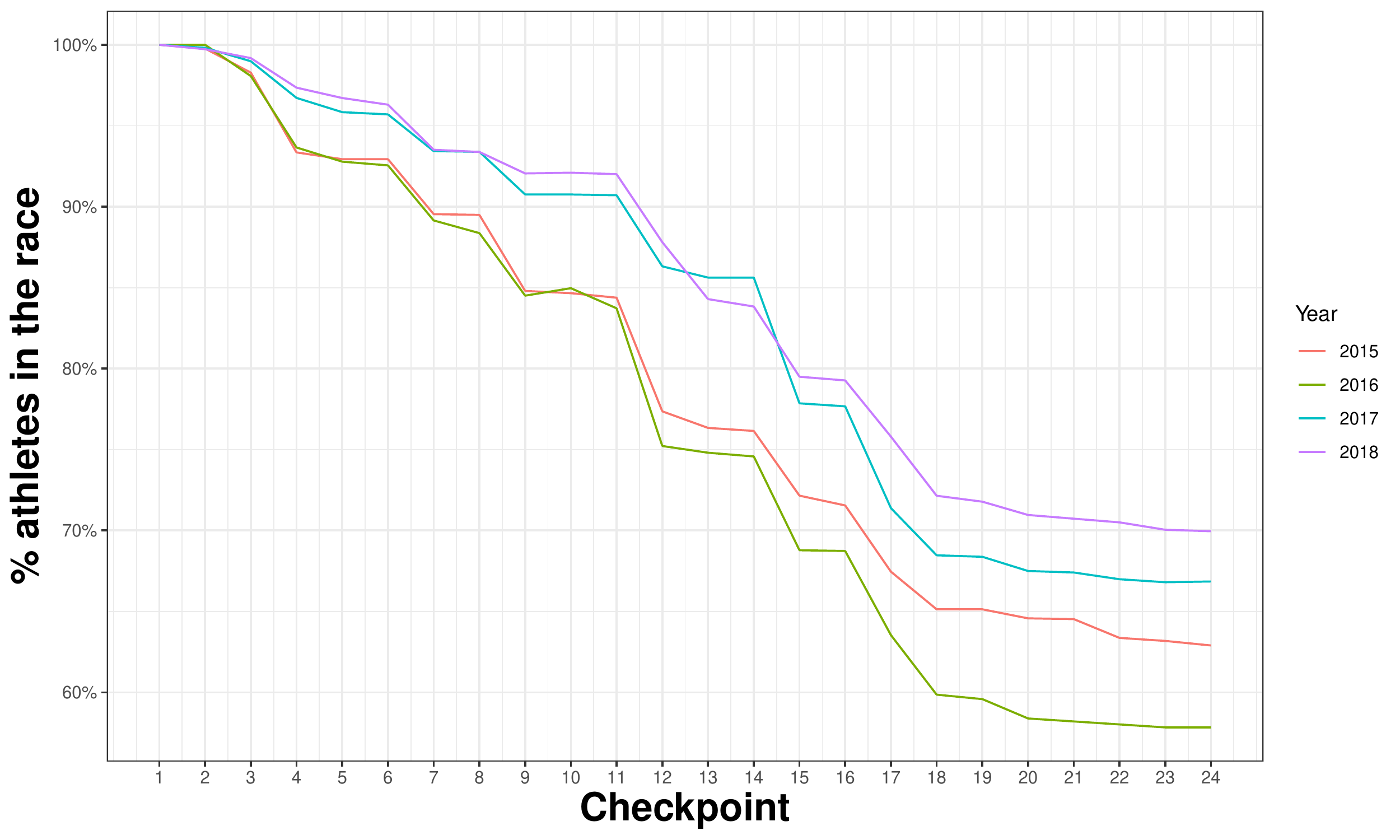}
  \caption{}
  \label{fig:dropout_rates}
\end{subfigure}%
\begin{subfigure}{.5\textwidth}
  \centering
  \includegraphics[width=\textwidth]{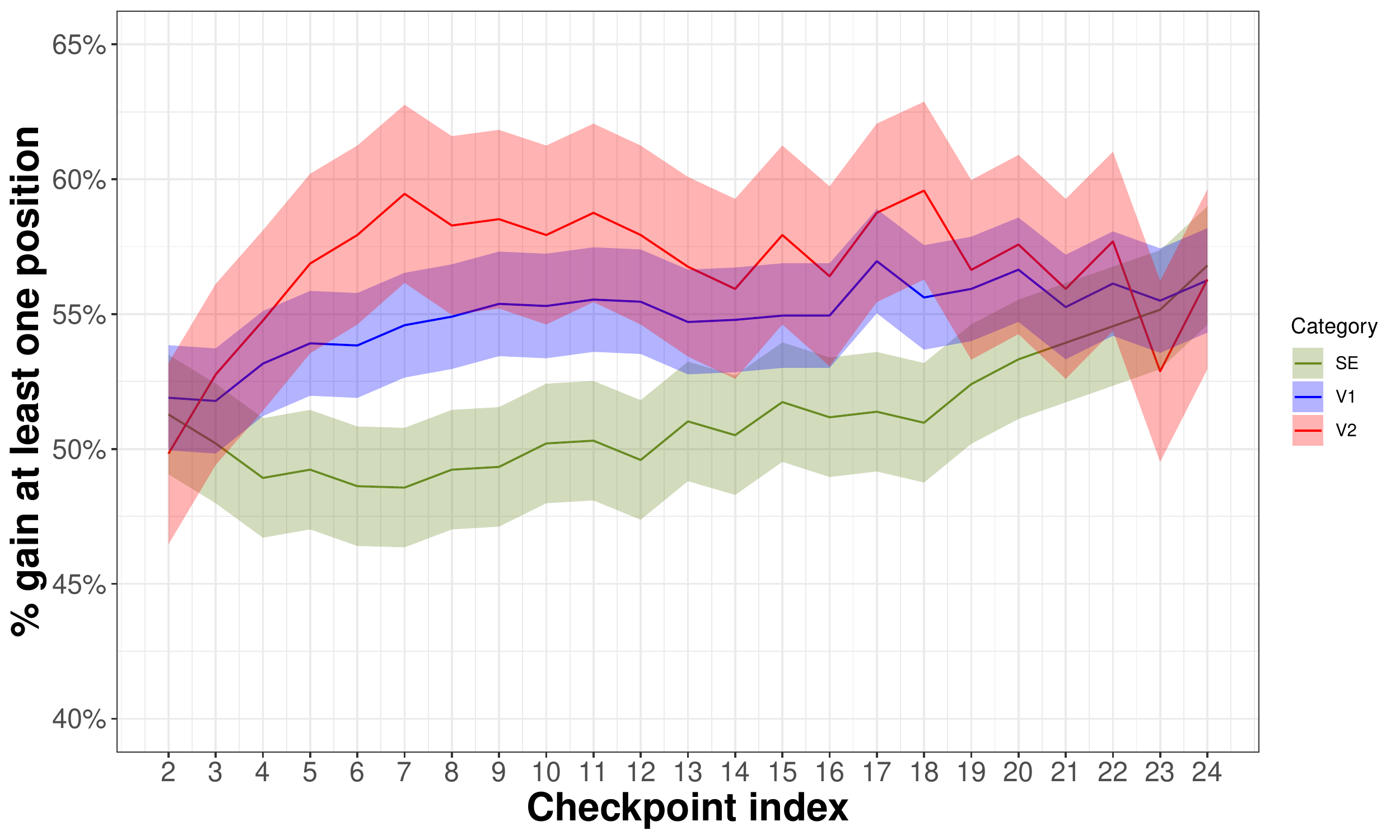}
  \caption{}
  \label{fig:gain_position_age}
 \end{subfigure}
\caption{(a) Cumulative proportion of athletes still in the race at the indicated checkpoints out of all athletes out of all athletes that started the race. (b) Fraction of runners, out of all the runners that completed the race, that gain at least one position in the total ranking, by category.} 
\end{figure}

How likely is it for runners to complete the race? Dropout rates of UTMB are consistently high across all editions considered, ranging between 30\% and 42\%. Interestingly, the editions with the highest dropout rates (2015 and 2016) were also characterized by bad weather conditions; it is very likely that weather played a key role in contributing to the observed differences. Not surprisingly, Figure~\ref{fig:dropout_rates} shows that dropouts are concentrated in a few small segments of the course: we observe large numbers of withdrawals between checkpoints 3-4 (Les Contamines-La Balme, only for 2015/16), 8-9 (Lac Combal-Mont Favre, only for 2015/16), 11-12 (Courmayeur-Bertone), 14-15 (Arnouvaz-Col Ferret), and 16-17 (La Fouly-Champex Lac). Although we do not have access to the exact locations and reasons behind the individual withdrawals, we can infer with some degree of certainty that most of them actually occur at checkpoints. Indeed, such large rates of dropouts may be explained by runners' fatigue or non serious injuries (e.g., blisters, ankle sprains), which do not prevent participants from walking and reaching the checkpoint. We note that all these checkpoints correspond to aid stations where both food and transportation provided by race organizers are available, that are easily accessible by supporters, and for which the distance to complete the race is still relatively long. Consequently, runners that are tired may have strong incentives to stop at these aid stations rather than on the trail or at checkpoints; this explanation is supported by the finding that a dropout is more likely if the runner has been decreasing pace, as reported in Section~\ref{sec:trap}. At the same time, a small fraction of the dropouts is often due to serious injuries; these events occur on the trail and require medical help. We also note that the total fraction of dropouts is slightly higher for women compared to men, with 42\% and 35\% respectively, and lower for younger runners (SE category) compared to older runners (V category), with 29\% and 39\%. This finding is interesting, as it shows that younger athletes complete the race more often despite likely being less experienced. Last, among the countries with at least 300 runners in these editions (France, Italy, Japan, Spain, UK, US), dropout rates appear to be highest for Italy with 40\% and lowest for Spain with 28\%. At the checkpoint level, we do not observe any significant difference in the locations of the dropouts across gender or age categories.\\

How does runners' pace vary throughout the race? Differently from the marathon setting, the uneven landscape of UTMB does not allow us to compare pace across different segments of the race. Indeed, each segment between checkpoints features unique distance, elevation, and difficulty, among several other characteristics; not all these variables are recorded in the data (e.g. difficulty) and consequently any analysis of pace across checkpoints requires assumptions on such variables. For this reason, instead of analysing (absolute) pace within groups, we focus on how (relative) pace compares across groups. In Figure~\ref{fig:gain_position_age}, we quantify these changes across age categories in terms of the fraction of runners that gain at least one position in the general ranking, out of all the runners that complete the race. A similar pattern can be observed analysing rankings based on partial times at checkpoints, but we do not report the related plots here. There are two notable patterns in Figure~\ref{fig:gain_position_age}. First, the fraction of runners that gain at least one position in the ranking is larger than 50\% across almost every checkpoint for all categories. This means that runners are more likely to gain, rather than to lose, positions in the ranking. Since such a gain needs to be offset by the loss of multiple positions by a lower fraction of runners, this implies that a small proportion of the runners may substantially decrease their pace. Second, younger runners (SE category) surprisingly appear to be more likely to gain at least one position towards the end of the race compared to older runners (V1 and V2 categories). We also notice that performance, in terms of the speed at checkpoints, of runners in the SE category increases more along the race than for the other two categories. We also investigate the relationships across genders and nationalities, but we do not find any significant pattern.



\section{TRAP framework}\label{sec:trap}
Our TRAP framework focuses on two random variables of interest. For each runner $i \in [n],$ at checkpoint $t,$ in a particular trail race with $T$ total checkpoints:
\begin{enumerate}
    \item $Y_{i,t} \in \mathbb{R}:$ passage time for runner $i$ at checkpoint $t$, and
    \item $D_{i,t} \in \{0, 1\}:$ whether runner $i$ drops out of the race at checkpoint $t$.
\end{enumerate}
Given $X_{i,t}$, a data structure representing the information about runner $i$ from the start of the race through checkpoint $t$, we introduce models for predicting three key quantities:
\begin{enumerate}
    \item $\mathbb{E}[Y_{i,t+1}| X_{i,t}]:$ expected passage time for runner $i$ at checkpoint $t+1$ given that runner $i$ has passed checkpoint $t$;
    \item $\left[Q_{.025}(Y_{i,t+1}| X_{i,t}), Q_{.975}(Y_{i,t+1}| X_{i,t})\right]: $ corresponding 95\% prediction interval for the checkpoint passage time; 
    \item $P(D_{i,t+1} | X_{i,t}): $ a probability mass function describing the likelihood that runner $i$ drops out of the race at checkpoint $t+1$ after passing checkpoint $t$.
\end{enumerate}
While our framework can be easily extended to predict runner's performance for multiple checkpoints ahead, in this work we focus on predictions for the only the next checkpoint. 
We proceed as follows. In Subsection~\ref{sec:features} we provide an overview of the different types of features used in modeling. In Subsection~\ref{sec:models} we describe the modeling approach. In Subsection~\ref{sec:validation} we describe models validation. In Section~\ref{sec:results} we summarise the results.

\subsection{Features}
\label{sec:features}

We constructed features for our data structure $X_{i,t}$ that can be clustered in three groups: based on checkpoint-level (Table~\ref{tab:ckptfeatures}), runner-level (Table~\ref{tab:runnerfeatures}), and lag information (Table~\ref{tab:laggedfeatures}).\\ Checkpoints characteristics, such as distance from starting point, altitude, food, and medical assistance, may vary across races and editions. To account for this heterogeneity, we manually encoded checkpoint-level information from tables and figures (e.g., Figure \ref{fig:utmb_profile}) available online. A summary of these features is listed in Table~\ref{tab:ckptfeatures}.\\

\begin{table}[]
\centering
\caption {Checkpoint-level features.}
\label{tab:ckptfeatures} 
\begin{tabular}{ll}
\hline
\textbf{Feature}        & \textbf{Description}                                                                       \\
\hline
InnerDist       & Distance from following checkpoint                                     \\
CumDist         & Cumulative distance from start of the race            \\
Altitude        & Altitude of checkpoint                                                    \\
CumulPlus       & Cumulative increase in altitude from start of the race                            \\
CumulMinus      & Cumulative decrease in altitude from start of the race                            \\
VarPlus         & Increase in altitude from last checkpoint                                     \\
VarMinus        & Decrease in altitude from last checkpoint                                     \\
Time\_barrier   & If there is maximum allowed time to reach the checkpoint                   \\
Drink           & If drinks are present at checkpoint                                  \\
Food            & If food is present at checkpoint                                    \\
Foodx2          & If warm meal is present at checkpoint                             \\
Bed             & If beds are present at checkpoint                           \\
Change\_clothes & If spare bag of clothes is present at checkpoint\\
Medical         & If first aid support is present at checkpoint \\
Bus             & If organization bus is present at checkpoint                            \\
WC              & If toilets are present at checkpoint\\
\hline
\end{tabular}
\end{table}

As described in Section~\ref{sec:utmb}, runner-level information was accessed from the UTMB and ITRA websites including demographics such as gender, nationality, and age category, together with information on past race history. Based on this set of information, we constructed predictors that range from for the runner's performance in previous races to features that characterize the types of races the runner took part in. The features that fall under this category are detailed in Table \ref{tab:runnerfeatures}.\\

\begin{table}[]
\centering
\caption {Runner-level features.} 
\label{tab:runnerfeatures} 
\begin{tabular}{ll}
\hline
\textbf{Feature}                    & \textbf{Description}                                                                                                                               \\
\hline
gender                        & Gender                                                                                                                           \\
nationality                & Nationality                                                                                                                      \\

category                   & Age category\\
n\_races                   & Number of races the runner has taken part in                                                                                               \\
mean\_rank\_perc           & Runner's average of $\frac{\textnormal{ranking}}{\textnormal{nr. participants}}$ in previous races                                                         \\
mean\_rank                 & Runner's average ranking considering all previous races                                                                                   \\
max\_rank\_perc            & Runner's max $\frac{\textnormal{ranking}}{\textnormal{
\# participants}}$ in previous races                                                                \\
min\_rank\_perc            & Runner's min $\frac{\textnormal{ranking}}{\textnormal{
\# participants}}$ in previous races                                                                \\
total\_elev                & Total elevation of previous races                                                                                          \\
mean\_elev                 & Average elevation of previous races                                                                                        \\
total\_dist                & Total distance of previous races                                                                                           \\
mean\_dist                 & Average distance of previous races                                                                                         \\
min\_dist                  & Min distance of previous races                                                                                             \\
max\_dist                  & Max distance of previous races                                                                                             \\
mean\_elev\_dist           & Average rate of elevation by distance of previous races                                                                       \\
max\_elev\_dist            & Maximal rate of elevation by distance of previous races                                                                           \\
min\_elev\_dist            & Minimum rate of elevation by distance of previous races                                                                           \\
n\_runners\_race           & Average number of competitors in previous races                                                                      \\
perc\_female\_race         & Average percentage of female runners in previous races                                                               \\
perc\_time\_overall        & Average of $\frac{\textnormal{Runner's time} - \textnormal{time of first position}}{\textnormal{time of last position} - \textnormal{time of first position}}$ in previous races \\
perc\_time\_first          & Average of $\frac{\textnormal{Runner's time} - \textnormal{time of first position}}{\textnormal{time of first position}}$ in previous races                         \\
perc\_time\_last           & Average of $\frac{\textnormal{Time of last finisher} - \textnormal{runner's time}}{\textnormal{time of last position}}$ in previous races                           \\
rank\_perc\_vs\_elev\_dist & Average of $\frac{\textnormal{Elevation}}{\textnormal{Distance}} \times \frac{\textnormal{Ranking}}{\textnormal{n\_runners\_race}}$ in previous races                                         \\
years\_activity            & Time (years) since the first ITRA race                                                \\
last\_year\_active         & Year of last ITRA race                                                  \\
time\_last\_race           & Time (months) since last ITRA race                                                                                                          \\
\hline
\end{tabular}
\end{table}

When predicting runner's performance throughout a race, one checkpoint at a time, there might exist a temporal dependency between the considered features and passage times. For example, one might expect that a runner underperforming in previous checkpoints will continue to underperform in upcoming checkpoints as well. For this reason, we include features based on lagged versions of the passage time: (lag 1) $Y_{i,t}$ and (lag 2) $Y_{i, t-1}$ when predicting the result at the next $t + 1$ checkpoint (both $Y_{i,t+1}$ and $D_{i,t+1}$). Additionally, we constructed several types of interaction variables between checkpoint-level information and the lagged passage time as presented in Table~\ref{tab:laggedfeatures}. We show in Section~\ref{sec:results} that the inclusion of the lagged features greatly contributes to the predictive power of the TRAP models.

\begin{table}[]
\centering
\caption {Lagged features including interactions between previous response time and an arbitrary upcoming checkpoint feature denoted by $x_{t+1}$.} 
\label{tab:laggedfeatures} 
\begin{tabular}{ll}
\hline
\textbf{Feature}        & \textbf{Description}                                                                       \\
\hline
lag\_1 and lag\_2    &  $Y_{i,t}$  and $Y_{i,t-1}$                                   \\
lag\_perc    &  $\frac{Y_{i,t}}{Y_{i,t} + Y_{i,t-1}}$ \\
lag\_1\_$x_{t+1}$ and  lag\_2\_$x_{t+1}$       &  $Y_{i,t} \cdot x_{t+1}$    and    $Y_{i, t-1} \cdot x_{t+1}$                                \\
lag\_perc\_$x_{t+1}$       &  $\frac{Y_{i,t}}{Y_{i,t} + Y_{i,t-1}} \cdot x_{t+1}$                                     \\
lag\_min\_1\_$x_{t+1}$ and   lag\_min\_2\_$x_{t+1}$    & $\frac{Y_{i,t} - \min_i Y_{t}}{\min_i Y_{t}} \cdot x_{t+1}$     and        $\frac{Y_{i,t-1} - \min_i Y_{t-1}}{\min_i Y_{t-1}} \cdot x_{t+1}$                         \\
\hline
\end{tabular}

\end{table}

\subsection{Models}
\label{sec:models}

\begin{table}[!b]
\centering
\caption{Comparison of considered models.}
\label{tab:models}
\scalebox{0.8}{
\begin{tabular}{|p{6cm} | p{3cm} | p{2cm} | p{2cm} |}
    \hline
    {\bf Model} & High-dimensions & Non-linear & Interactions \\
    \hline
    Intercept-only (baseline) & & &  \\    
    \hline
    LASSO & \checkmark & &  \\    
    \hline
    XGBoost & \checkmark & \checkmark & \checkmark \\    
    \hline
    Random forest & \checkmark & \checkmark & \checkmark  \\   
    \hline
\end{tabular}}
\vspace{1em}
\end{table}

Given the potential features considered for the data structure $X_{i,t}$, the models in the TRAP framework must be able to address the following conditions: (1) be high-dimensional, (2) account for non-linear relationships, and (3) account for interactions. We considered many different types of variables in Subsection~\ref{sec:features}, resulting in a high-dimensional problem with the potential for non-linear relationships and interactions, e.g. types of runners perform better at certain types of checkpoints. Thus, we considered models that capture these aspects of the data. \\

For modeling both the expected passage time $\mathbb{E}[Y_{i,t+1}| X_{i,t}]$ and probability of dropping out $P(D_{i,t+1} | X_{i,t})$, we established a \textit{baseline} intercept-only model without any of the features described in Section \ref{sec:features}. The intercept-only model serves as a reference point for the model evaluation process described in Subsection~\ref{sec:validation}.\\

Next, we considered the LASSO regression model \citep{tibshirani94} to handle the high-dimensional nature of the problem, while providing fast and interpretable results. We use the \texttt{glmnet} implementation \citep{glmnet} in the \texttt{R} programming language \citep{R17}. For the expected passage time model we used the default mean-squared error (MSE) measure to choose the regularization parameter with cross-validation (CV), while for the probability of dropping out model we used logistic LASSO regression with the area under the curve (AUC) measure. For both models we chose the one standard error regularization penalty.\\

In order to account for potential non-linear relationships and interactions, we additionally explored the usage of tree-based models. We considered both random forests \citep{Breiman01}, via the \texttt{ranger} package \citep{ranger17}, and gradient boosted trees via XGBoost in \texttt{R} \citep{Chen16, xgboost}. Both tree-based models were trained using the default MSE loss for the expected passage time model. For the probability of dropping out, both the LASSO and XGBoost were trained to maximize the AUC. For random forests, we train probability forests as described in \citet{malley2012probability}. We found one hundred trees for both random forests and XGBoost models to yield the best results, where the boosted trees were limited to a maximum depth of three, based on CV results.\\

We concurrently selected the modeling framework for the prediction intervals based on which expected passage time model was selected according to the criteria described in Subsection~\ref{sec:validation}. Based on the results in Section~\ref{sec:results}, we generated 95\% prediction intervals with random forests using the quantile regression framework described by \citet{meinshausen2006quantile}, as implemented in the \texttt{ranger} package via the \texttt{quantreg} option. Table~\ref{tab:models} summarizes the different models considered. Of course with more explicit feature engineering beyond what is listed in Subsection~\ref{sec:features} the LASSO can also account for non-linear interactions, but for our purposes we compare its performance to the tree-based models as the simple, linear model reference.

\subsection{Model Validation}
\label{sec:validation}

To ensure that our selected models perform well in more than one race, we used leave-one-year-out (LOYO) CV (e.g., train on all runner and race checkpoints in years 2015 through 2017 and test on 2018). To evaluate the models we used root mean-squared error (RMSE) and area under the curve (AUC) for the expected passage time and probability of dropping out models, respectively. We evaluated the models based on two criteria: (1) overall performance measures, and (2) the performance measures across all checkpoints to avoid model overfitting to certain checkpoints.\\

In addition to evaluating the types of models in Section~\ref{sec:models}, we also evaluated the types of features presented in Subsection~\ref{sec:features}, separated in four groups:
\begin{enumerate}
    \item Checkpoint,
    \item Checkpoint + Runner,
    \item Checkpoint + Runner + Lag 1,
    \item Checkpoint + Runner + Lag 1 \& 2.
\end{enumerate}
Due to the inclusion of the lag information from two checkpoints prior, we only compared both model and feature performance starting at the third checkpoint in each year.

\section{Results}\label{sec:results}
This section walks through the results and analysis of the models in our TRAP framework.

\subsection{Expected passage time model comparison}
\label{sec:regression-comparison}

\begin{figure}[t]
\begin{subfigure}{.5\textwidth}
  \centering
  \includegraphics[width=\textwidth]{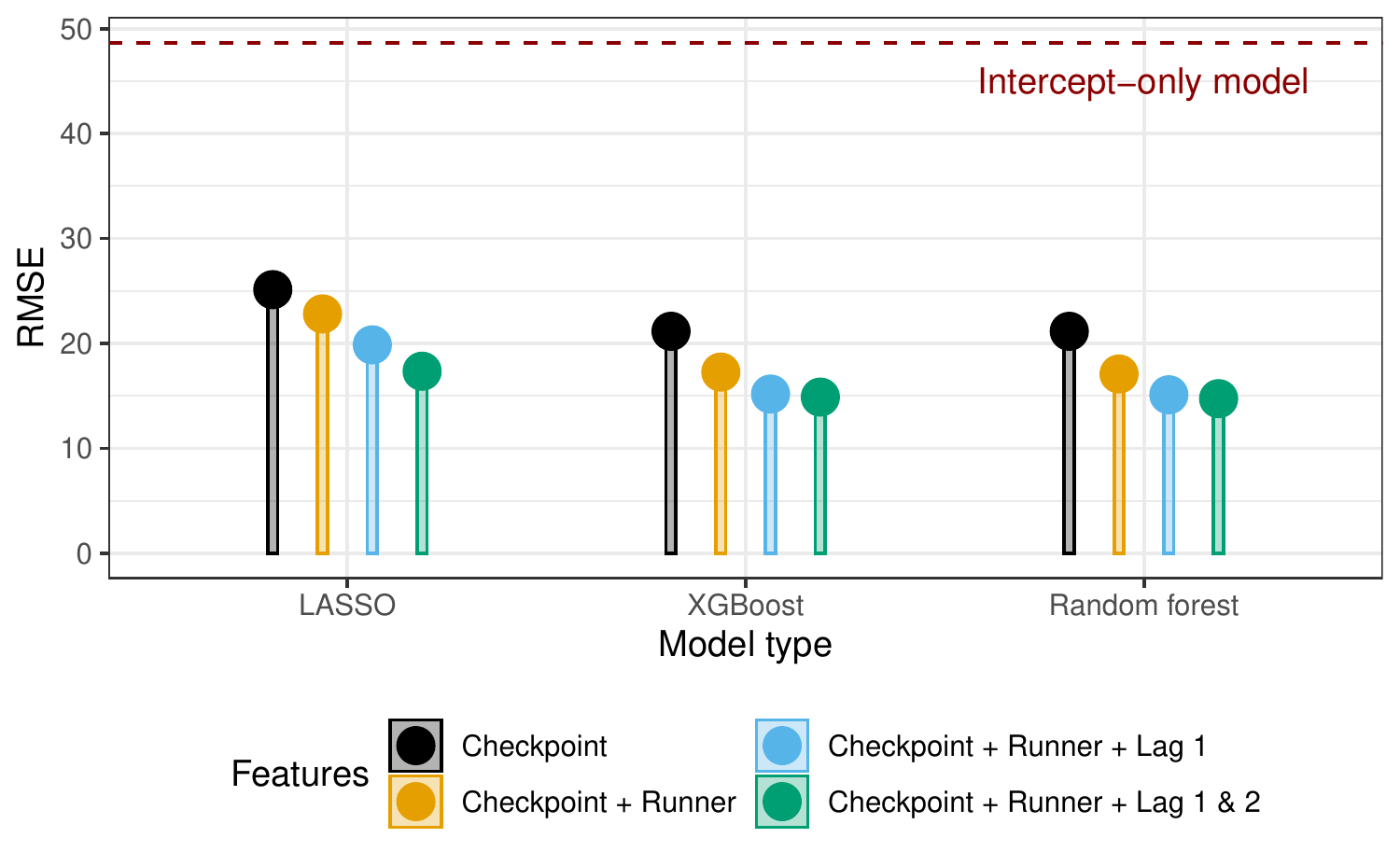}
  \caption{}
  \label{fig:loyo-rmse-overall}
\end{subfigure}%
\begin{subfigure}{.5\textwidth}
  \centering
  \includegraphics[width=\textwidth]{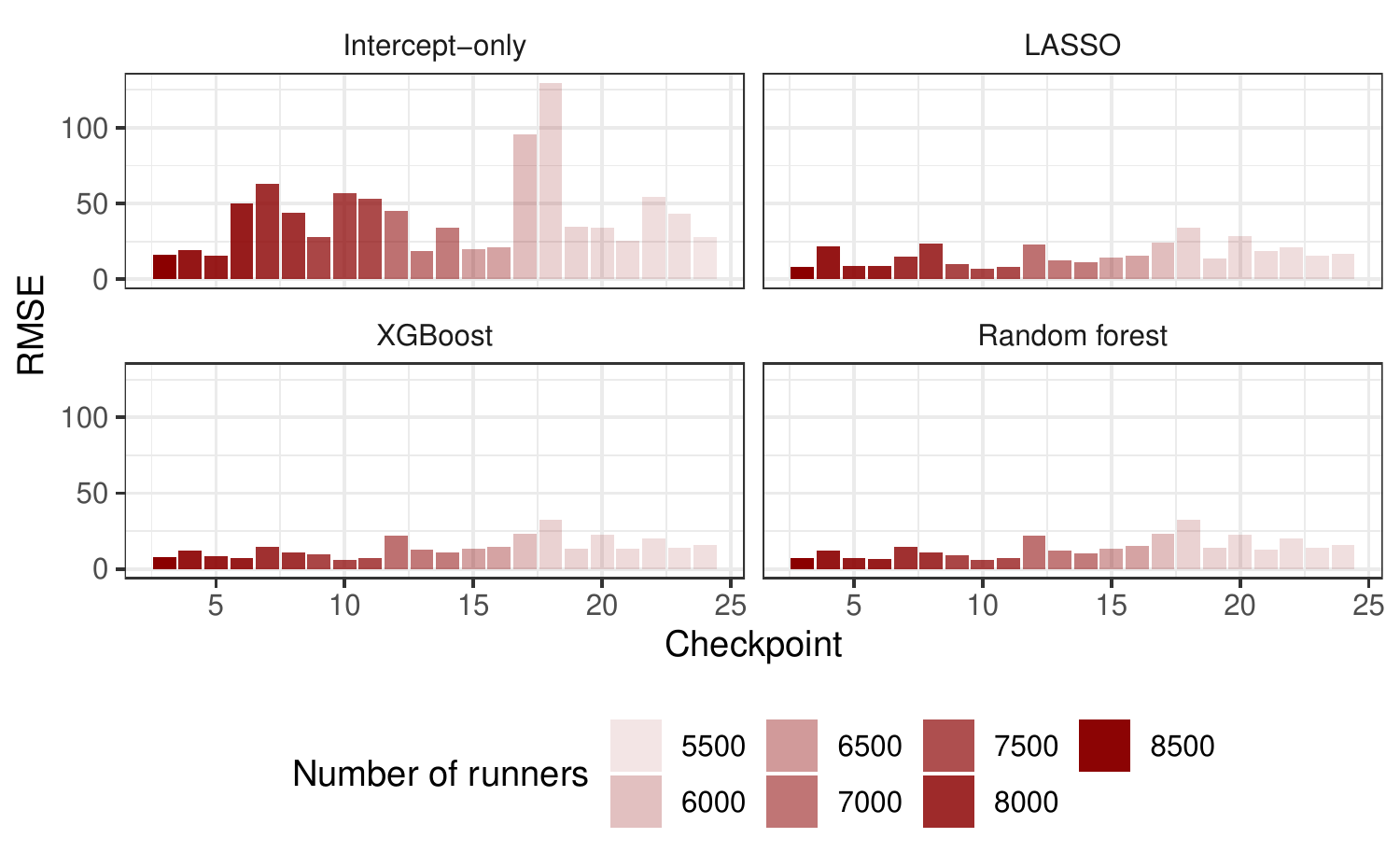}
  \caption{}
  \label{fig:loyo-rmse-ckp}
 \end{subfigure}
\caption{(a) LOYO CV RMSE by type of model and features considered for expected passage time. Dashed line indicates intercept-only model results. (b) LOYO CV RMSE by checkpoint and model type for expected passage time.} 
\end{figure}

Figure \ref{fig:loyo-rmse-overall} displays the overall LOYO CV RMSE for each type of expected passage time model presented in Section \ref{sec:models} along with the features considered in Section \ref{sec:features}. All possible model-feature combinations largely outperform the intercept-only model, but the tree-based approaches appear to display a better performance than the LASSO for each set of features. In particular, the drop in RMSE given by the inclusion of both checkpoint- and runner-level information is larger for the tree-based models, indicating the advantage from accounting for potential interactions in the data. The addition of both sets of lag variables also generate an improvement in model performance, with a smaller improvement from including features that were based on two checkpoints prior. \\

Figure \ref{fig:loyo-rmse-ckp} displays the checkpoint level LOYO CV RMSE for each model type using the full set of features (Checkpoint + Runner + Lag 1 \& 2) in addition to the reference baseline model. The three models vastly outperform the intercept-only approach and unsurprisingly they all result in higher RMSE values for the checkpoints closer to the end of the race likely due to fewer runners remaining. Based on the results seen in Figures~\ref{fig:loyo-rmse-overall} and~\ref{fig:loyo-rmse-ckp}, we determined that the random forest and XGBoost models yield comparable results, which are better than those of the baseline and LASSO models overall and across checkpoints. However, due to its simple extension, we use the quantile regression forest implementation of random forests in the \texttt{ranger} package for generating the 95\% prediction intervals in Section \ref{sec:intervals}.

\subsection{Probability of dropping out model comparison}
\label{sec:classification-comparison}

\begin{figure}[t]
\begin{subfigure}{.5\textwidth}
  \centering
  \includegraphics[width=\textwidth]{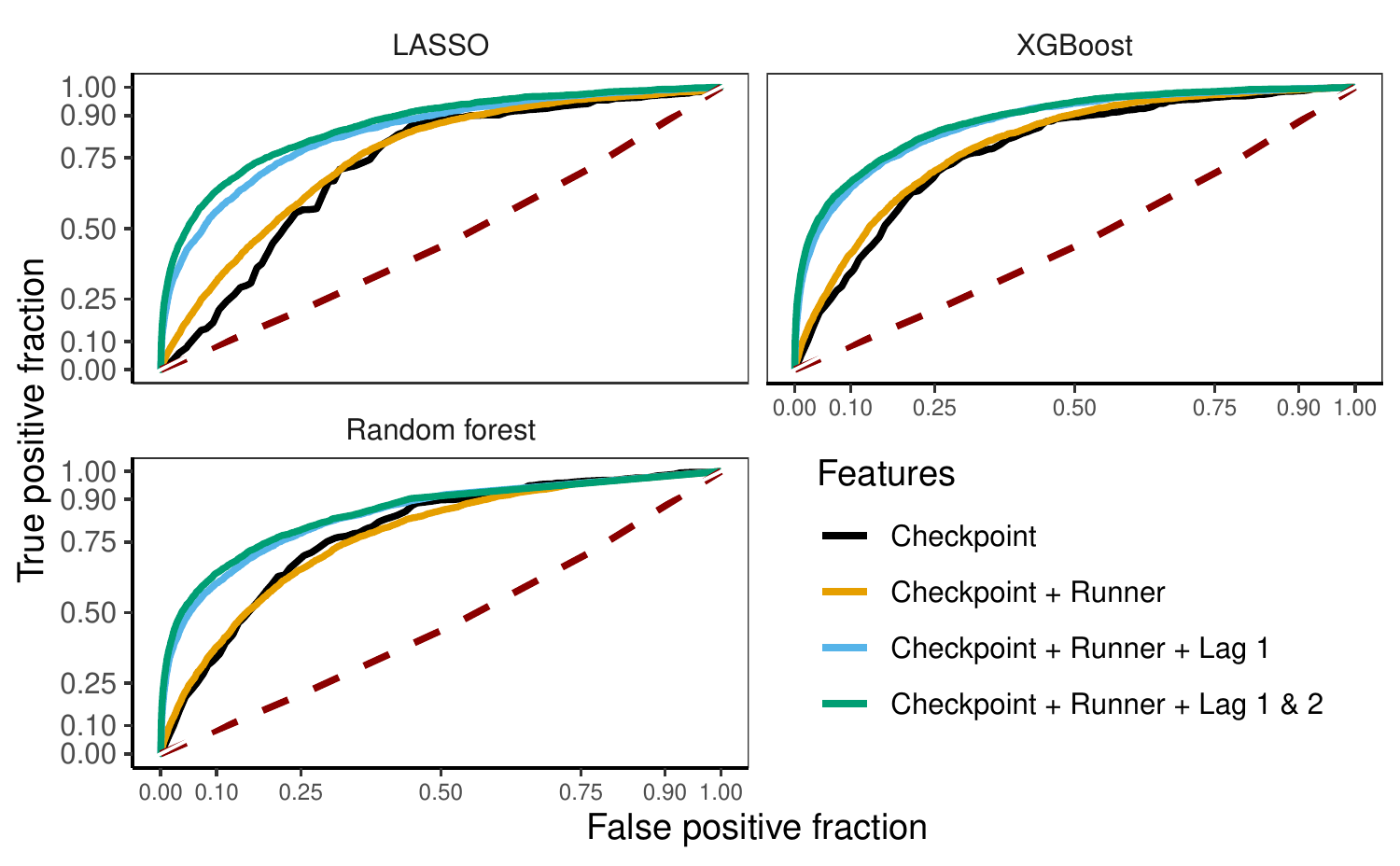}
  \caption{}
  \label{fig:loyo-roc-curves}
\end{subfigure}%
\begin{subfigure}{.5\textwidth}
  \centering
  \includegraphics[width=\textwidth]{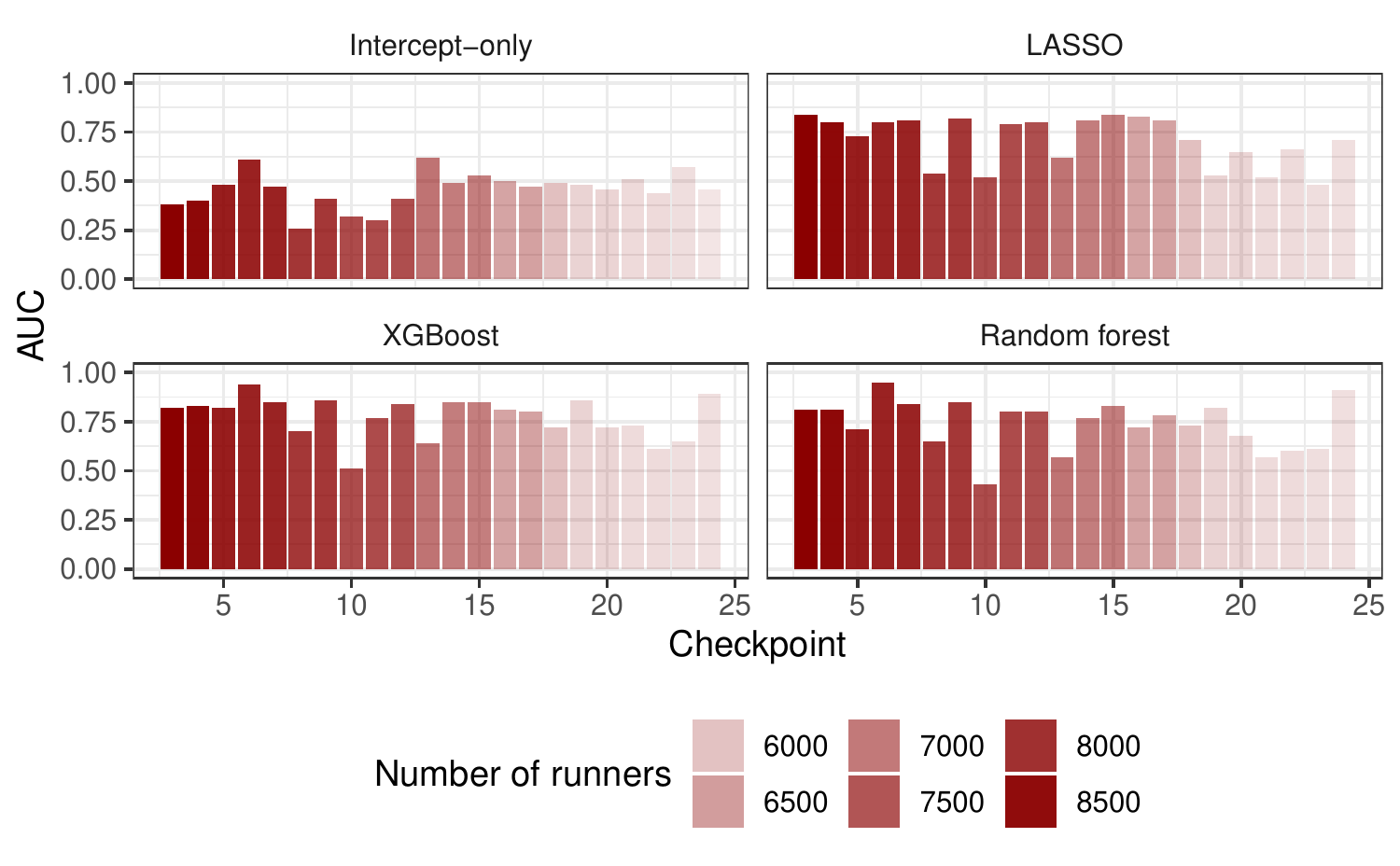}
  \caption{}
  \label{fig:loyo-auc-ckp}
 \end{subfigure}
\caption{(a) LOYO CV ROC curve by type of model and features considered for probability of dropping out model. Dashed line indicates intercept-only model results. (b) LOYO CV AUC by checkpoint and model type for probability of dropping out.} 
\end{figure}

Figure \ref{fig:loyo-roc-curves} displays the LOYO CV receiver operating characteristic (ROC) curves for each model-feature combination considered for the probability of dropping out model. Table \ref{tab:auc-results} displays the corresponding overall AUC for the LOYO CV results. In comparison to the expected passage time results presented in Subsection \ref{sec:regression-comparison}, the inclusion of the runner-level information in addition to the checkpoints does not appear to lead to an improvement in model performance. However, as before, all combinations outperform the intercept-only baseline and inclusion of the lag features (back to two checkpoints prior) leads to the best holdout performance across the different models considered.\\

\begin{table}[!b]
\caption{LOWO CV AUC by model type and set of features. Intercept-only (baseline) displayed LOWO CV AUC equal to 0.46.}
\label{tab:auc-results}
\centering
\centering
\scalebox{0.8}{
\begin{tabular}{|p{4cm} | p{3cm} | p{2cm} | p{2cm} | p{2cm} |}
    \hline
    {\bf Model} & Checkpoint & + Runner & + Lag 1 & + Lag 2 \\
    \hline
    LASSO & 0.72 & 0.75 & 0.84 & 0.86\\    
    \hline
    XGBoost & 0.78 & 0.80 & 0.87 & 0.88 \\    
    \hline
    Random forest & 0.78 & 0.77 & 0.85 & 0.86 \\  
    \hline
\end{tabular}}
\end{table}

Figure \ref{fig:loyo-auc-ckp} displays the checkpoint level LOYO CV AUC values for each model type using the full set of features in comparison to the baseline intercept-only model. Again, all three models outperform the intercept-only and display a decay in performance for later checkpoints. We determine that the best performance is achieved by the XGBoost model based on the overall holdout AUC measure and performance across all checkpoints. We do not observe any notable differences between the holdout years for both the expected passage time and probability of dropping out models.

\subsection{Analysis of Feature Importance}
\label{sec:importance}

\begin{figure}
    \begin{subfigure}{.5\textwidth}
  \centering
  \includegraphics[width=\textwidth]{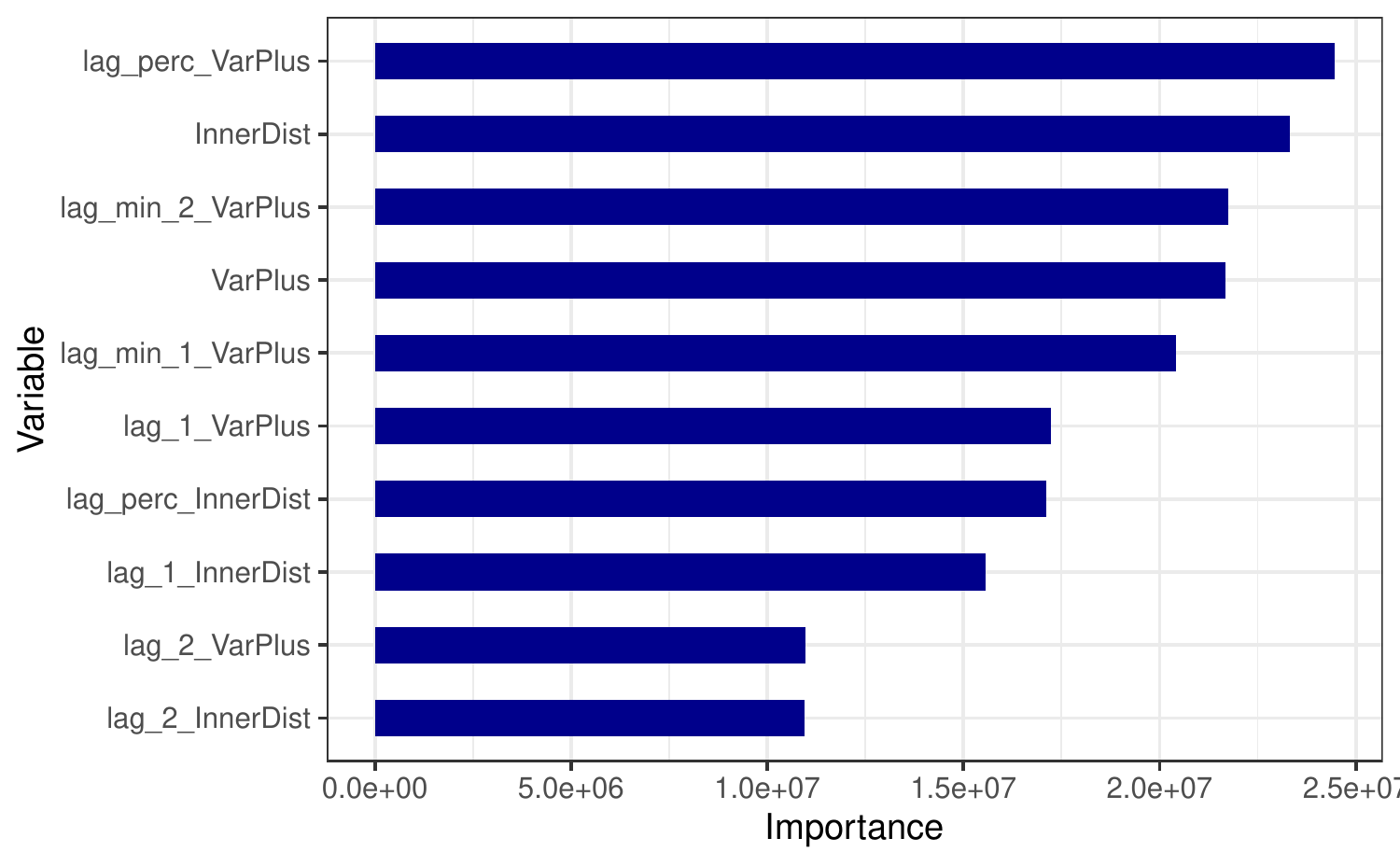}
  \caption{}
  \label{fig:var-imp-plot1}
\end{subfigure}%
\begin{subfigure}{.5\textwidth}
  \centering
  \includegraphics[width=\textwidth]{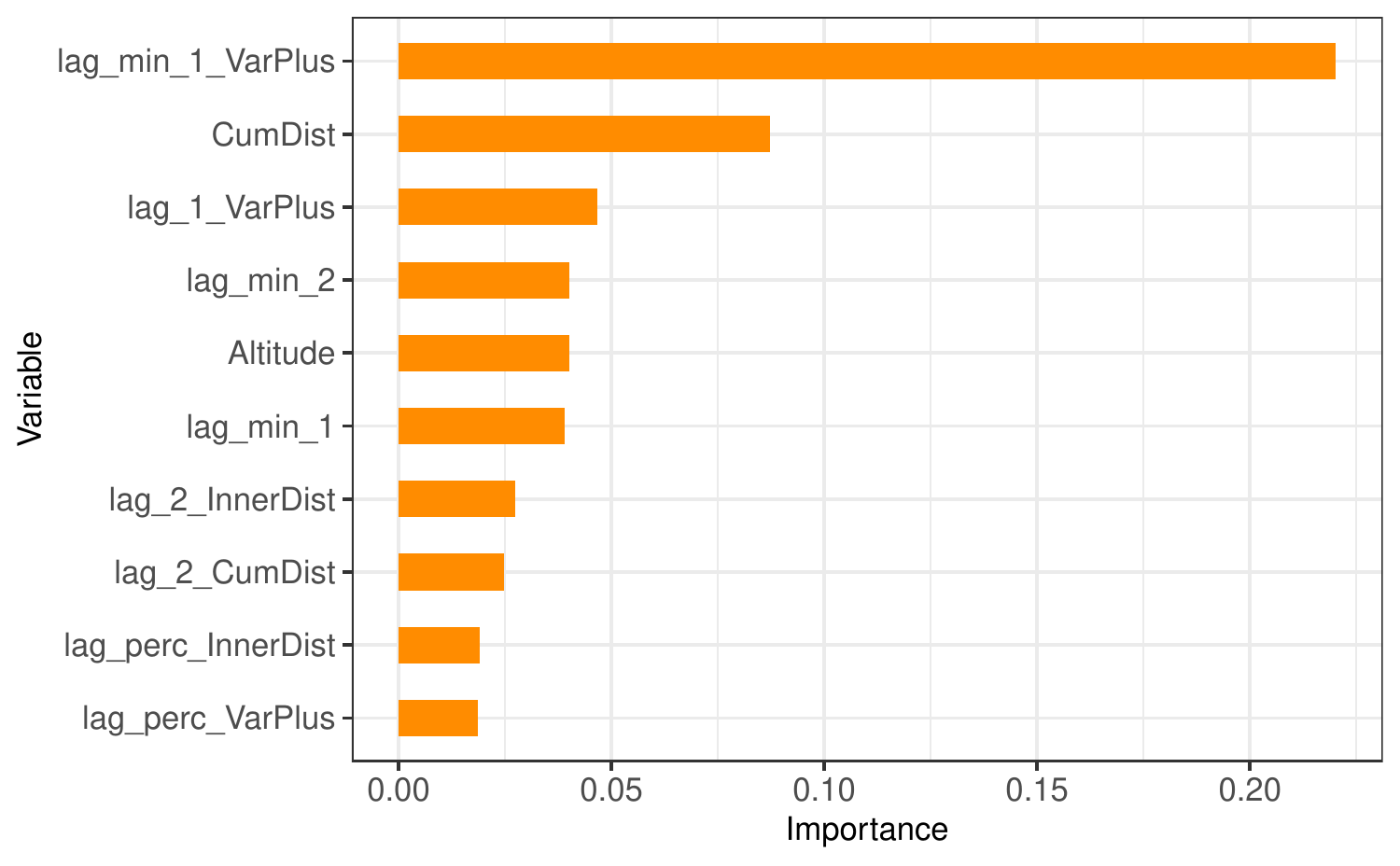}
  \caption{}
  \label{fig:var-imp-plot2}
 \end{subfigure}
 \caption{Top ten variables by importance for (a) random forest expected passage time model and (b) XGBoost probability of dropping out model.}
    \label{fig:var-imp-plots}
\end{figure}

To provide more context with regards to the models we find to yield the best performance, Figures \ref{fig:var-imp-plot1} and \ref{fig:var-imp-plot2} display the top ten features by importance for the random forest expected passage time and XGBoost probability of dropping out models respectively. Unsurprisingly, the most important variable for predicting the expected passage time at the checkpoint is the interaction between gain in elevation from the last checkpoint and the ratio of time passed from the passage at the last checkpoint and from the second to last checkpoint. The second most important variable is the distance from the last checkpoint.  Eight of the top ten listed features are interactions between lagged and checkpoint-level variables. Interestingly, the most important variable for predicting the probability of dropping out is an interaction between the increase in altitude and a relative 1-lagged response, indicating an increase in the likelihood of dropping out if the runner was slower in the previous checkpoint when compared to the first runner that reached the checkpoint. This variable is followed naturally by the cumulative distance traveled,  reflecting the increase in the likelihood of dropping out as the runner continues in the race.

\subsection{Prediction Interval Examples}
\label{sec:intervals}

\begin{figure}
    \centering
    \includegraphics[width = 0.8\textwidth]{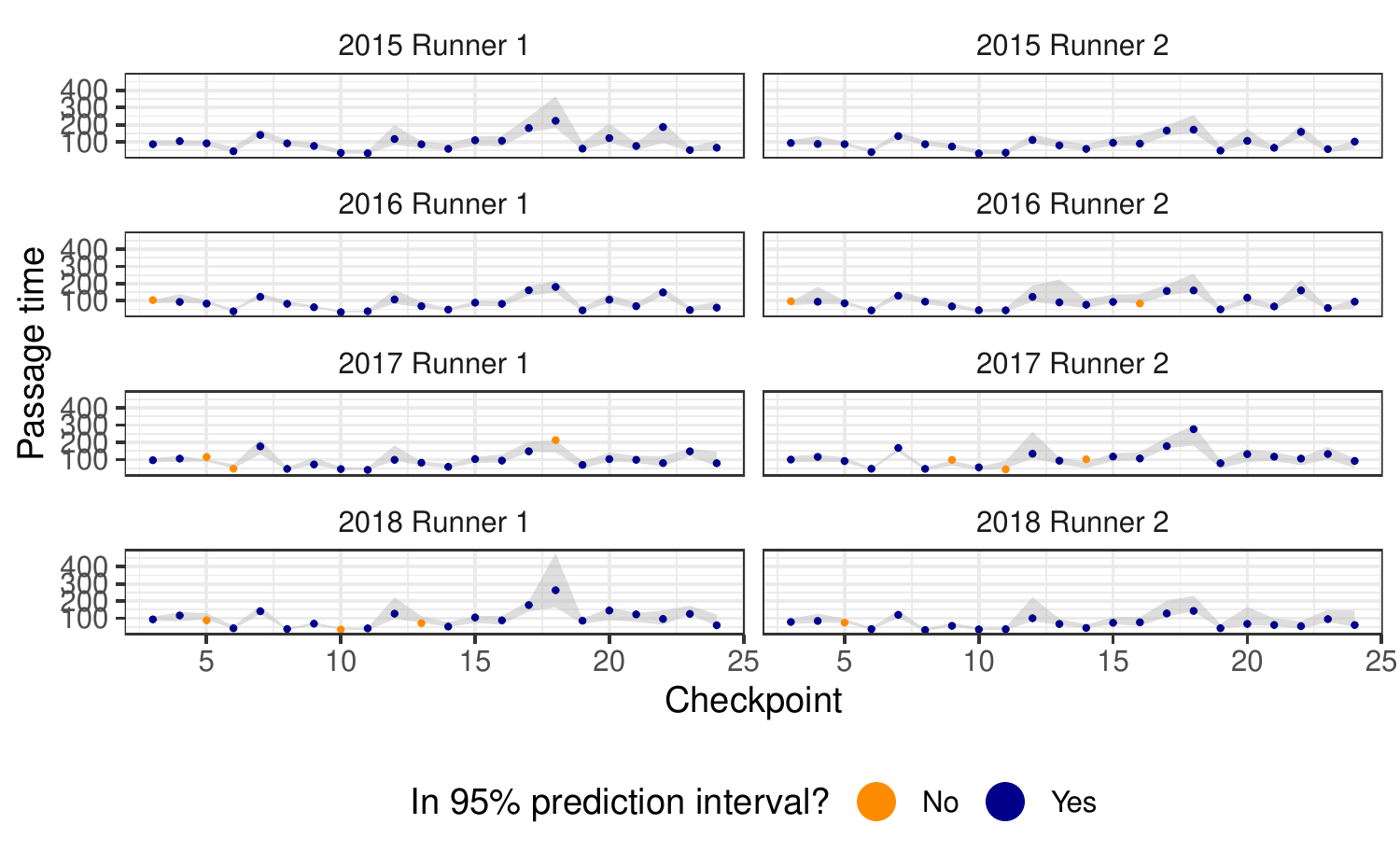}
    \caption{Examples of performance for two randomly selected runners from each edition of UTMB in the data set, from 2015 to 2018, that completed the race. Each runner's actual passage time is displayed with black points along with 95\% prediction intervals indicated with gray bands.}
    \label{fig:example-runner-pi}
\end{figure}

Figure \ref{fig:example-runner-pi} displays an example of 95\% prediction intervals for a series of randomly selected runners that completed the entire race for each available year. These prediction intervals were generated out-of-sample similar to the LOYO CV. For example, the prediction intervals for the 2018 race are based on the quantile regression forest trained on years 2015-2017, with an overall out-of-sample coverage of 91.4\% across all runners, races, and checkpoints. Each point corresponds to the observed passage time by the runner, while the gray bands indicate the 95\% prediction intervals. The color of the point indicates whether or not the the point is inside the prediction interval. While we notice examples of passage times outside of the prediction intervals, we see that they are not drastically off. Additionally, we note differences between the pairs of runners for each year, with certain checkpoints varying in the size of the interval. For example, we see a clear difference in prediction interval and then performance for the two runners in 2018 following checkpoint 16. We also explored the usage of features constructed based on these intervals (e.g., observed passage time within 95\% prediction interval for previous checkpoint, cumulative in or out of interval, etc.), but did not find an improvement in model performance relative to the usage of the lagged information.

\section{Discussion}\label{sec:discussion}




In this work we have presented a predictive framework, TRAP, built on the race history of runners accessed from ITRA. Our contributions serve two distinct but complementary purposes. On the one hand, we have made available to the public (available on GitHub  \texttt{https://github.com/ricfog/TRAP\_data}) a data set containing race- and runner-level information recorded on the ITRA website. We envision the simplified access to the this data set, together with our seminal work, to encourage research on trail running. On the other hand, TRAP is informative in the phase of course design and assessment of runners' performance happening before and during the race respectively. Design entails decisions regarding the location of checkpoints and aid stations, evaluation of the segments and of race difficulty, and consequent placement of race volunteers and medical staff. We have shown that information about the runners' past race history can be incorporated into the design phase through the TRAP framework prior to the start of the race. At the same time, assessment of performance throughout the race plays another fundamental role.
The likelihood of dropout and prediction intervals inform race organizers of runners that are experiencing difficulties and may need medical help. Expected passage times and prediction intervals are also useful to supporters and race volunteers to understand the time range in which runners will reach the following checkpoint. 
TRAP represents a transferable and scalable framework: the model needs not be retrained for new races and the accuracy of its predictions increases with data. \\

Despite the successful results presented in Section~\ref{sec:trap}, our framework suffers from limitations related to the matching of runners within and between data sets. First of all, matching of runners' names is a problematic issue. Indeed, in the ITRA data set, the same name could refer to multiple runners. Since we used names and nationalities to aggregate race records, this could have caused measurement errors in the race history for a tiny fraction of runners. This issue could be partially solved by matching runners also on the age category. Alternatively, scraping runner-level information from the runners' ITRA profiles, instead of constructing it from race-level data as we did, would completely solve this issue. We did not implement this strategy due to the large size of the ITRA data set. Matching of runners between UTMB and ITRA data sets is also complicated by names mispellings. The fraction of participants that we could not match and consequently dropped from our analysis of UTMB is likely explained by names mispellings in one of the two data sets. Given the set of information currently available, we leave this problem for future work with record linkage methodology. Last, we recognize that ITRA data might be only partially informative of the runners' past experience, particularly for runners from countries such as the US, where only a fraction of all races is recorded in ITRA.\\

This work opens several future directions of research. On the ITRA side, scraping runner-level data from the runners' webpages would provide information about age and ITRA performance scores. It would also help researchers circumnavigate the aforementioned problem of duplicate names. Analyses of the ITRA data set could focus on runners' performance over time or differences across nationalities and age categories.
On the race-level side, as it is the case for UTMB, an interesting line of research is represented by the analysis of runners performance, leveraging data sources that provide fine-grid information. For example, the addition of features such as weather conditions and details on the terrain would likely improve the performance of the models. One could also investigate performance using Strava and Garmin Connect data, which provide a (almost) time continuous stream of information on runners' position and pace. Such data would allow researchers to examine runners' pacing strategies in a manner similar to sports with player-tracking data, e.g. evaluating continuous-time movements in basketball \citep{Sicilia19} and American football \citep{yurko2019going}. An extension of the TRAP methodology could incorporate this filtration $\mathcal{F}(X_{i,t})$ of information via recurrent neural networks, and provide improved, continuous-time insights into the segments difficulties. Another improvement of TRAP would be possible if data regarding the reason behind the dropout was available, such as why the runner withdrew from the race. The disentanglement of dropouts caused by serious medical issues from other reasons is of central interest to race organizers and medical staff. While predictions of these events would likely suffer from a high degree of uncertainty, they might still represent an interesting target to consider. 


\newpage
\bibliographystyle{DeGruyter}
\bibliography{references}

\begin{thebibliography}{43}
\newcommand{\enquote}[1]{``#1''}
\providecommand{\natexlab}[1]{#1}
\providecommand{\url}[1]{\texttt{#1}}
\providecommand{\urlprefix}{URL }

\bibitem[{Angermeier({2018})}]{anger15}
Angermeier, A. ({2018}): \enquote{Mountaineers and endurance athletes are
  stoked by trail running,}
  \url{https://www.ispo.com/en/trends/trail-running-gets-popular-among-endurance-athletes-and-mountaineers},
  Last accessed on 2020-06-19.

\bibitem[{Bartolucci and Murphy(2015)}]{bartolucci2015finite}
Bartolucci, F. and T.~B. Murphy (2015): \enquote{A finite mixture latent
  trajectory model for modeling ultrarunners’ behavior in a 24-hour race,}
  \emph{Journal of Quantitative Analysis in Sports}, 11, 193--203.

\bibitem[{Breiman(2001)}]{Breiman01}
Breiman, L. (2001): \enquote{Random forests,} \emph{Machine Learning}, 45,
  5--32, \urlprefix\url{https://doi.org/10.1023/A:1010933404324}.

\bibitem[{Burgess(2020)}]{burgess15}
Burgess, M. (2020): \enquote{How ultrarunners are pushing the human body beyond
  all limits,} \emph{Wired},
  \url{https://www.wired.co.uk/article/ultra-marathon-running-uk-tom-evans},
  Last accessed on 2020-06-19.

\bibitem[{{Cambridge Dictionary}(2020)}]{camb}
{Cambridge Dictionary} (2020): \enquote{Definition of trail running,}
  \urlprefix\url{https://dictionary.cambridge.org/us/dictionary/english/trail-running},
  last accessed on 2020-06-18.

\bibitem[{Chen and Guestrin(2016)}]{Chen16}
Chen, T. and C.~Guestrin (2016): \enquote{Xgboost: A scalable tree boosting
  system,} in \emph{Proceedings of the 22Nd ACM SIGKDD International Conference
  on Knowledge Discovery and Data Mining}, KDD '16, New York, NY, USA: ACM,
  785--794, \urlprefix\url{http://doi.acm.org/10.1145/2939672.2939785}.

\bibitem[{Chen et~al.(2019)Chen, He, Benesty, Khotilovich, Tang, Cho, Chen,
  Mitchell, Cano, Zhou, Li, Xie, Lin, Geng, and Li}]{xgboost}
Chen, T., T.~He, M.~Benesty, V.~Khotilovich, Y.~Tang, H.~Cho, K.~Chen,
  R.~Mitchell, I.~Cano, T.~Zhou, M.~Li, J.~Xie, M.~Lin, Y.~Geng, and Y.~Li
  (2019): \emph{xgboost: Extreme Gradient Boosting},
  \urlprefix\url{https://CRAN.R-project.org/package=xgboost}, r package version
  0.81.0.1.

\bibitem[{Coast et~al.(2004)Coast, Blevins, and Wilson}]{coast2004gender}
Coast, J.~R., J.~S. Blevins, and B.~A. Wilson (2004): \enquote{Do gender
  differences in running performance disappear with distance?} \emph{Canadian
  Journal of Applied Physiology}, 29, 139--145.

\bibitem[{Cuk et~al.(2020)Cuk, Nikolaidis, and Knechtle}]{cuk2020sex}
Cuk, I., P.~T. Nikolaidis, and B.~Knechtle (2020): \enquote{Sex differences in
  pacing during half-marathon and marathon race,} \emph{Research in Sports
  Medicine}, 28, 111--120.

\bibitem[{Deaner et~al.(2015)Deaner, Carter, Joyner, and
  Hunter}]{deaner2015men}
Deaner, R.~O., R.~E. Carter, M.~J. Joyner, and S.~K. Hunter (2015):
  \enquote{Men are more likely than women to slow in the marathon,}
  \emph{Medicine and science in sports and exercise}, 47, 607.

\bibitem[{Ely et~al.(2007)Ely, Cheuvront, Roberts, and Montain}]{ely2007impact}
Ely, M.~R., S.~N. Cheuvront, W.~O. Roberts, and S.~J. Montain (2007):
  \enquote{Impact of weather on marathon-running performance.} \emph{Medicine
  and Science in Sports and Exercise}, 39, 487--493.

\bibitem[{Friedman et~al.(2010)Friedman, Hastie, and Tibshirani}]{glmnet}
Friedman, J., T.~Hastie, and R.~Tibshirani (2010): \enquote{Regularization
  paths for generalized linear models via coordinate descent,} \emph{Journal of
  Statistical Software}, 33, 1--22,
  \urlprefix\url{http://www.jstatsoft.org/v33/i01/}.

\bibitem[{Friedrich et~al.(2014)Friedrich, R{\"u}st, Rosemann, Knechtle,
  Barandun, Lepers, and Knechtle}]{friedrich2014comparison}
Friedrich, M., C.~A. R{\"u}st, T.~Rosemann, P.~Knechtle, U.~Barandun,
  R.~Lepers, and B.~Knechtle (2014): \enquote{A comparison of anthropometric
  and training characteristics between female and male half-marathoners and the
  relationship to race time,} \emph{Asian journal of sports medicine}, 5, 10.

\bibitem[{Haney~Jr and Mercer(2011)}]{haney2011description}
Haney~Jr, T.~A. and J.~A. Mercer (2011): \enquote{A description of variability
  of pacing in marathon distance running,} \emph{International Journal of
  Exercise Science}, 4, 133.

\bibitem[{Hoffman(2014)}]{hoffman2014pacing}
Hoffman, M.~D. (2014): \enquote{Pacing by winners of a 161-km mountain
  ultramarathon,} \emph{International journal of sports physiology and
  performance}, 9, 1054--1056.

\bibitem[{Hubble and Zhao(2016)}]{hubble2016gender}
Hubble, C. and J.~Zhao (2016): \enquote{Gender differences in marathon pacing
  and performance prediction,} \emph{Journal of Sports Analytics}, 2, 19--36.

\bibitem[{{International Association of Athletics Federations}(2015)}]{iaaf15}
{International Association of Athletics Federations} (2015): \enquote{Iaaf
  congress, beijing, china, 19 august 2015 – day 1 notes,}
  \url{https://www.worldathletics.org/news/press-release/iaaf-congress-beijing-2015},
  Last accessed on 2020-06-19.

\bibitem[{Ives(2015)}]{ives15}
Ives, M. (2015): \enquote{Running in the wild,} \emph{The New York Times},
  \url{https://www.nytimes.com/2015/09/01/travel/trail-marathons-running-in-the-wild.html},
  Last accessed on 2020-06-18.

\bibitem[{Keogh et~al.(2019)Keogh, Smyth, Caulfield, Lawlor, Berndsen, and
  Doherty}]{keogh2019prediction}
Keogh, A., B.~Smyth, B.~Caulfield, A.~Lawlor, J.~Berndsen, and C.~Doherty
  (2019): \enquote{Prediction equations for marathon performance: A systematic
  review,} \emph{International Journal of Sports Physiology and Performance},
  14, 1159--1169.

\bibitem[{Kerherv{\'e} et~al.(2016)Kerherv{\'e}, Cole-Hunter, Wiegand, and
  Solomon}]{kerherve2016pacing}
Kerherv{\'e}, H.~A., T.~Cole-Hunter, A.~N. Wiegand, and C.~Solomon (2016):
  \enquote{Pacing during an ultramarathon running event in hilly terrain,}
  \emph{PeerJ}, 4, e2591.

\bibitem[{Knechtle et~al.(2014)Knechtle, Barandun, Knechtle, Zingg, Rosemann,
  and R{\"u}st}]{knechtle2014prediction}
Knechtle, B., U.~Barandun, P.~Knechtle, M.~A. Zingg, T.~Rosemann, and C.~A.
  R{\"u}st (2014): \enquote{Prediction of half-marathon race time in
  recreational female and male runners,} \emph{Springerplus}, 3, 248.

\bibitem[{Knechtle et~al.(2011)Knechtle, Knechtle, Barandun, Rosemann, and
  Lepers}]{knechtle2011predictor}
Knechtle, B., P.~Knechtle, U.~Barandun, T.~Rosemann, and R.~Lepers (2011):
  \enquote{Predictor variables for half marathon race time in recreational
  female runners,} \emph{Clinics}, 66, 287--291.

\bibitem[{Knechtle et~al.(2015)Knechtle, Rosemann, Zingg, Stiefel, and
  R{\"u}st}]{knechtle2015pacing}
Knechtle, B., T.~Rosemann, M.~A. Zingg, M.~Stiefel, and C.~A. R{\"u}st (2015):
  \enquote{Pacing strategy in male elite and age group 100 km
  ultra-marathoners,} \emph{Open Access Journal of Sports Medicine}, 6, 71.

\bibitem[{Krawczyk and Wilamowski(2017)}]{krawczyk2017we}
Krawczyk, M. and M.~Wilamowski (2017): \enquote{Are we all overconfident in the
  long run? evidence from one million marathon participants,} \emph{Journal of
  Behavioral Decision Making}, 30, 719--730.

\bibitem[{Lambert et~al.(2004)Lambert, Dugas, Kirkman, Mokone, and
  Waldeck}]{lambert2004changes}
Lambert, M.~I., J.~P. Dugas, M.~C. Kirkman, G.~G. Mokone, and M.~R. Waldeck
  (2004): \enquote{Changes in running speeds in a 100 km ultra-marathon race,}
  \emph{Journal of sports science \& medicine}, 3, 167.

\bibitem[{Malley et~al.(2012)Malley, Kruppa, Dasgupta, Malley, and
  Ziegler}]{malley2012probability}
Malley, J.~D., J.~Kruppa, A.~Dasgupta, K.~G. Malley, and A.~Ziegler (2012):
  \enquote{Probability machines: Consistent probability estimation using
  nonparametric learning machines,} \emph{Methods of information in medicine},
  51, 74--81.

\bibitem[{March et~al.(2011)March, Vanderburgh, Titlebaum, and
  Hoops}]{march2011age}
March, D.~S., P.~M. Vanderburgh, P.~J. Titlebaum, and M.~L. Hoops (2011):
  \enquote{Age, sex, and finish time as determinants of pacing in the
  marathon,} \emph{The Journal of Strength \& Conditioning Research}, 25,
  386--391.

\bibitem[{Meinshausen(2006)}]{meinshausen2006quantile}
Meinshausen, N. (2006): \enquote{Quantile regression forests,} \emph{Journal of
  Machine Learning Research}, 7, 983--999.

\bibitem[{Nikolaidis et~al.(2019)Nikolaidis, Cuk, Rosemann, and
  Knechtle}]{nikolaidis2019performance}
Nikolaidis, P.~T., I.~Cuk, T.~Rosemann, and B.~Knechtle (2019):
  \enquote{Performance and pacing of age groups in half-marathon and marathon,}
  \emph{International journal of environmental research and public health}, 16,
  1777.

\bibitem[{Nikolaidis and Knechtle(2017)}]{nikolaidis2017effect}
Nikolaidis, P.~T. and B.~Knechtle (2017): \enquote{Effect of age and
  performance on pacing of marathon runners,} \emph{Open access journal of
  sports medicine}, 8, 171.

\bibitem[{Peter et~al.(2014)Peter, Rust, Knechtle, Rosemann, and
  Lepers}]{peter2014sex}
Peter, L., C.~A. Rust, B.~Knechtle, T.~Rosemann, and R.~Lepers (2014):
  \enquote{Sex differences in 24-hour ultra-marathon performance-a
  retrospective data analysis from 1977 to 2012,} \emph{Clinics}, 69, 38--46.

\bibitem[{{R Core Team}(2017)}]{R17}
{R Core Team} (2017): \emph{R: A Language and Environment for Statistical
  Computing}, R Foundation for Statistical Computing, Vienna, Austria,
  \urlprefix\url{https://www.R-project.org/}.

\bibitem[{R{\"u}st et~al.(2011)R{\"u}st, Knechtle, Knechtle, Barandun, Lepers,
  and Rosemann}]{rust2011predictor}
R{\"u}st, C.~A., B.~Knechtle, P.~Knechtle, U.~Barandun, R.~Lepers, and
  T.~Rosemann (2011): \enquote{Predictor variables for a half marathon race
  time in recreational male runners,} \emph{Open access journal of sports
  medicine}, 2, 113.

\bibitem[{R{\"u}st et~al.(2015)R{\"u}st, Rosemann, Zingg, and
  Knechtle}]{rust2015non}
R{\"u}st, C.~A., T.~Rosemann, M.~A. Zingg, and B.~Knechtle (2015): \enquote{Do
  non-elite older runners slow down more than younger runners in a 100 km
  ultra-marathon?} \emph{BMC Sports Science, Medicine and Rehabilitation}, 7,
  1.

\bibitem[{Saiidi({2020})}]{saiidi15}
Saiidi, U. ({2020}): \enquote{Think running 26.2 miles is tough? these runners
  are going 62 miles,} \emph{Cnbc},
  \url{https://www.cnbc.com/2020/03/06/sports-ultramarathons-are-becoming-very-popular-in-asia.html},
  Last accessed on 2020-06-19.

\bibitem[{Santos-Lozano et~al.(2014)Santos-Lozano, Collado, Foster, Lucia, and
  Garatachea}]{santos2014influence}
Santos-Lozano, A., P.~Collado, C.~Foster, A.~Lucia, and N.~Garatachea (2014):
  \enquote{Influence of sex and level on marathon pacing strategy. insights
  from the new york city race,} \emph{International journal of sports
  medicine}, 35, 933--938.

\bibitem[{Sicilia et~al.(2019)Sicilia, Pelechrinis, and Goldsberry}]{Sicilia19}
Sicilia, A., K.~Pelechrinis, and K.~Goldsberry (2019): \enquote{Deephoops:
  Evaluating micro-actions in basketball using deep feature representations of
  spatio-temporal data,} \emph{ACM SIGKDD}.

\bibitem[{Smyth(2018)}]{smyth2018fast}
Smyth, B. (2018): \enquote{Fast starters and slow finishers: a large-scale data
  analysis of pacing at the beginning and end of the marathon for recreational
  runners,} \emph{Journal of Sports Analytics}, 4, 229--242.

\bibitem[{Tibshirani(1996)}]{tibshirani94}
Tibshirani, R. (1996): \enquote{Regression shrinkage and selection via the
  lasso,} \emph{Journal of the Royal Statistical Society. Series B
  (Methodological)}, 58, 267--288,
  \urlprefix\url{http://www.jstor.org/stable/2346178}.

\bibitem[{Trubee et~al.(2014)Trubee, Vanderburgh, Diestelkamp, and
  Jackson}]{trubee2014effects}
Trubee, N.~W., P.~M. Vanderburgh, W.~S. Diestelkamp, and K.~J. Jackson (2014):
  \enquote{Effects of heat stress and sex on pacing in marathon runners,}
  \emph{The Journal of Strength \& Conditioning Research}, 28, 1673--1678.

\bibitem[{Wright and Ziegler(2017)}]{ranger17}
Wright, M.~N. and A.~Ziegler (2017): \enquote{{ranger}: A fast implementation
  of random forests for high dimensional data in {C++} and {R},} \emph{Journal
  of Statistical Software}, 77, 1--17.

\bibitem[{Yurko et~al.(2020)Yurko, Matano, Richardson, Granered, Pospisil,
  Pelechrinis, and Ventura}]{yurko2019going}
Yurko, R., F.~Matano, L.~F. Richardson, N.~Granered, T.~Pospisil,
  K.~Pelechrinis, and S.~L. Ventura (2020): \enquote{Going deep: models for
  continuous-time within-play valuation of game outcomes in american football
  with tracking data,} \emph{Journal of Quantitative Analysis in Sports}.

\bibitem[{Zingg et~al.(2014)Zingg, Karner-Rezek, Rosemann, Knechtle, Lepers,
  and R{\"u}st}]{zingg2014will}
Zingg, M.~A., K.~Karner-Rezek, T.~Rosemann, B.~Knechtle, R.~Lepers, and C.~A.
  R{\"u}st (2014): \enquote{Will women outrun men in ultra-marathon road races
  from 50 km to 1,000 km?} \emph{Springerplus}, 3, 97.

\end{thebibliography}

\pagebreak
\setcounter{page}{1}
\begin{center}
\textbf{\large Supplementary Materials for:\\ \vspace{2cm} ``TRAP: A Predictive Framework for Trail Running Assessment of Performance"}
\end{center}
\pagebreak

\setcounter{equation}{0}
\setcounter{figure}{0}
\setcounter{table}{0}
\setcounter{section}{0}
\makeatletter
\renewcommand{\theequation}{S\arabic{equation}}
\renewcommand{\thefigure}{S\arabic{figure}}
\renewcommand{\bibnumfmt}[1]{[S#1]}
\renewcommand{\citenumfont}[1]{S#1}



\section{ITRA background}
\label{sec:scrapitra}

The International Trail Running Association (ITRA) is the world's largest trail running organization. Funded in July 2013, ITRA is nonprofit and its central mission is the promotion of trail running as a sport. Its rankings are considered the benchmark for the evaluation of races (e.g. difficulty) and athletes (performance) in the trail running community. However, ITRA is not the only organization in this space. Notably, UltraSignup is another organization with a large community in North America. ITRA and UltraSignup differ in their ranking systems, in the services offered to their members, and in the way memberships are managed. For instance, ITRA requires race organizers to pay for membership through a flat fee whose cost ranges from 100 to 550 Euros per year depending on the size of the organization but independently of the number of races. Through the membership, ITRA provides an evaluation of the race's difficulty and of runners' performances. All data is recorded in their database; for timing services, many races in the ITRA circuit use the LiveTrail software described in Section~\ref{sec:related}. UltraSignup, instead, charges a fee for each runner but provides a free timing software. 
We focus on ITRA because, besides being the largest organization, its circuit of races also includes the case-study of this paper. 
The main features of ITRA are: (1) the evaluation of the race's difficulty and the system of ``points''; (2) the ``performance'', i.e. ranking, system.\\

(1) Each race in the ITRA circuit is assigned ``ITRA points'', a score proportional to its difficulty. The assignment of the number of points is based on a measure called ``km-effort'', which is the sum of distance travelled (in km) plus one point for every 100 m in vertical gain, minus penalties proportional to the number of refreshment points and laps. The km-effort difficulty of the race is then converted into points through a standard table, whose categories correspond to the ones used in Figure~\ref{fig:nraces_ITRA}. 
The maximum number of points is capped at 6. For instance, UTMB is approximately 171 km long with 10000 m of elevation gain and no laps, therefore the km-effort measure is around $171+100=271$ which translates into 6 ITRA points.
Every race finisher gains the corresponding points and can use them to qualify for future -- and harder -- races. Indeed, most races above 50 km require participants to have a minimum number of points to qualify. As an example, in 2019 UTMB runners needed 10 ITRA points obtained in maximum two races only to enter the lottery; this means that they needed either both a $\geq 210$ and a $115-154$ km-effort race or two $155-209$ km-effort races. The points can only be gained through races in the ITRA circuit. \\

(2) ITRA has adopted a scale, called the ``ITRA performance index'', that allows for comparison of athletes' performances across races. Every race finisher gains an ITRA score, which is a function of the km-effort, the technical difficulty of the race, and the hypothetical best finish time in that race.\footnote{Differently from the computation of the points, ITRA has not openly released an explicit equation for the score.}
Multiple scores are then aggregated into the athlete's general and categories indexes; the categories breakdown follows the legend of Figure~\ref{fig:nraces_ITRA}, but in terms of km-effort. Both indexes are calculated as the mean of the 5 highest ITRA scores weighted decreasingly by time and up to 36 months before.  The index ranges from 0 to 1000 points, with elites and amateurs scoring 800 and around 500/600 respectively. Despite the drawback of being proprietary knowledge, the ITRA index has become the benchmark for the evaluation of athletes' performances in trail running. \\

\section{Exploratory analysis of ITRA data}
\label{sec:itradata}

We are interested in both races results and evaluations, and athletes' performance and scores. We accessed the results of all races available on the ITRA website from 2010 to the end of 2018\footnote{Data was accessed via the \texttt{ScrapITRA} Python package: \texttt{https://github.com/ricfog/ScrapITRA}}. The downloaded data set contains over 2.9 million individual results\footnote{We only observe the runner's name (1.07 million unique names in our data set), so we can not reliably match runners across races. Indeed, two unique individuals may have the same name, sex, and nationality.} in more than 15 thousand different races. Figures~\ref{fig:nraces_ITRA} and~\ref{fig:country_races_ITRA} are based on this data set. We now present a few other findings from our exploratory data analysis on this ITRA data set. We describe, in sequence, results regarding races and runners. While the following data analysis is by no means extensive, it serves the purpose of presenting the main features of the data set. \\

\begin{figure}[t]
\begin{subfigure}{.5\textwidth}
  \centering
  \includegraphics[width=\textwidth]{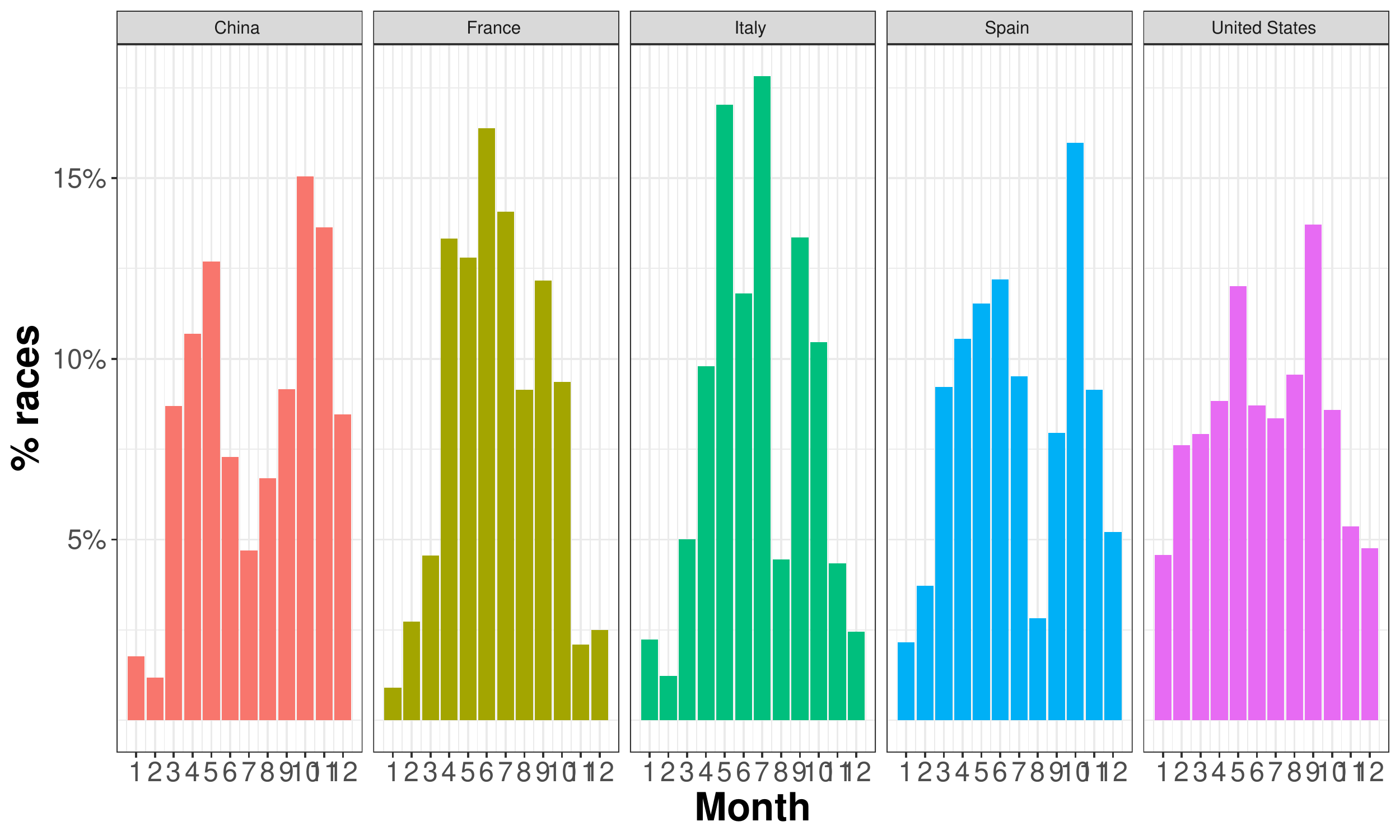}
  \caption{}
  \label{fig:races_by_month}
\end{subfigure}%
\begin{subfigure}{.5\textwidth}
  \centering
  \includegraphics[width=\textwidth]{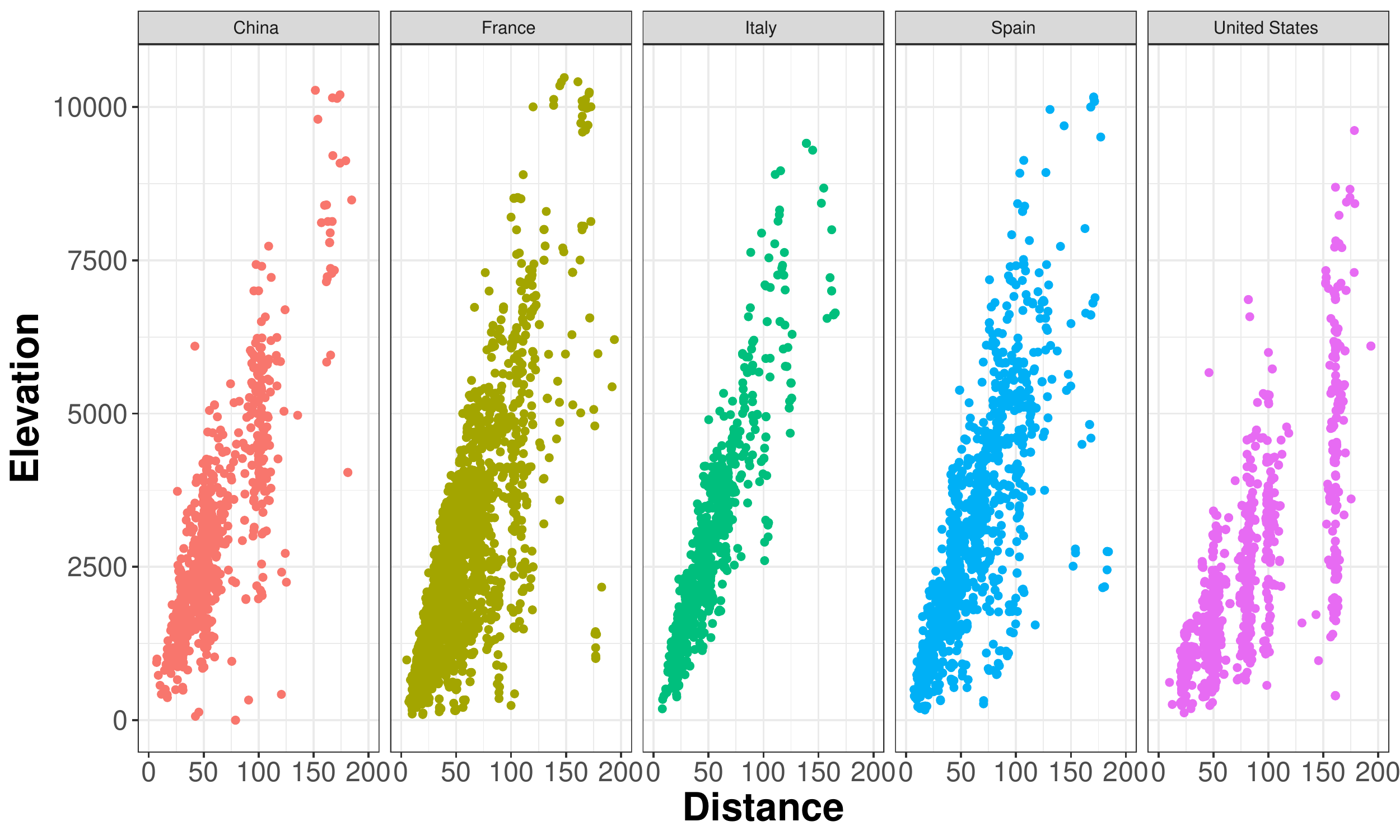}
  \caption{}
  \label{fig:dist_elev_by_country}
\end{subfigure}
\caption{Data is taken from the ITRA data set for the five countries in the data set with at least 500 races each in the years 2010-2018. (a) Histogram distribution of races across months. The ticks 1-12 on the horizontal axis  correspond to the months January-December. The vertical axis represents the ratio of the number of races in a given month and in the corresponding year respectively. (b) Elevation and distance for races in the indicated countries. Each point corresponds to a race.}
\end{figure}

We start with the information on races. We find a relationship between the race's distance, elevation, and period of the year. Figure~\ref{fig:races_by_month} shows the histogram distribution of races across months for the years from 2010 until 2018 for the five countries in the data set that have at least 500 races each. Every bar represents the ratio of the numbers of races in a given month and in the corresponding year. We notice some degree of heterogeneity among the distributions. Among these five countries, only in France, Italy, and Spain the majority of races is held in the months from April to October. In Figure~\ref{fig:dist_elev_by_country}, which shows the relationship between distance and elevation for ITRA races, we observe that competitions in these countries feature high elevation gains and therefore are probably held on mountain trails.
For instance, races in Italy feature, on average, 56 m/km of elevation gain thanks to the fact that they are mainly held in the Alps. Consequently, the lower number of races in winter is likely due to factors such as low temperatures and snow on trails. Interestingly, for these three countries there appears to be a sizeable decrease in the number of races in the month of August; this effect may be explained by the high temperatures or by workers' vacations.
The distributions of races across months for the US and China have different shapes. In the US, races are more evenly distributed across spring, summer, and fall. Figure~\ref{fig:dist_elev_by_country} shows that races in this country are typically longer but have relatively less elevation gain, with only 32 m/km. The absence of steep ascents, and consequently the presence of trails that are more flat, may allow for an easier organization of races in all seasons. Differently, in China most races are held either in spring or fall, and only a few of them are in the summer. 
Our explanation of these existing patterns between period and elevation gains focuses on the geographical location rather than on organizers' or runners' preferences. However, this explanation can not be ruled out. Indeed, in Figure~\ref{fig:dist_elev_by_country} we notice that both the US and China have only a few races with distance between 100 and 160km, which may reflect runners' preferences.\\

\begin{figure}[t]
\begin{subfigure}{.5\textwidth}
  \centering
  \includegraphics[width=\textwidth]{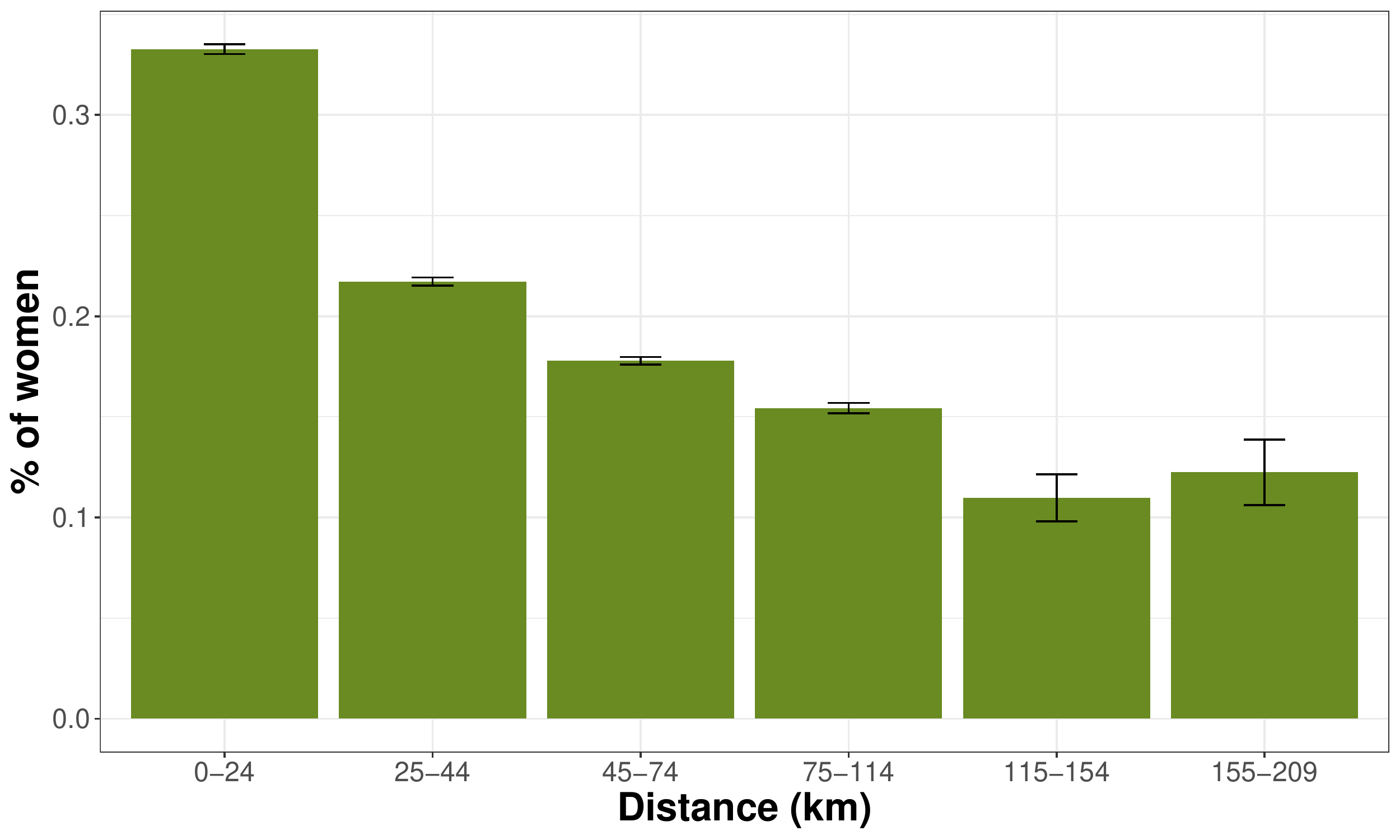}
  \caption{}
  \label{fig:women_by_distance}
\end{subfigure}%
\begin{subfigure}{.5\textwidth}
  \centering
  \includegraphics[width=\textwidth]{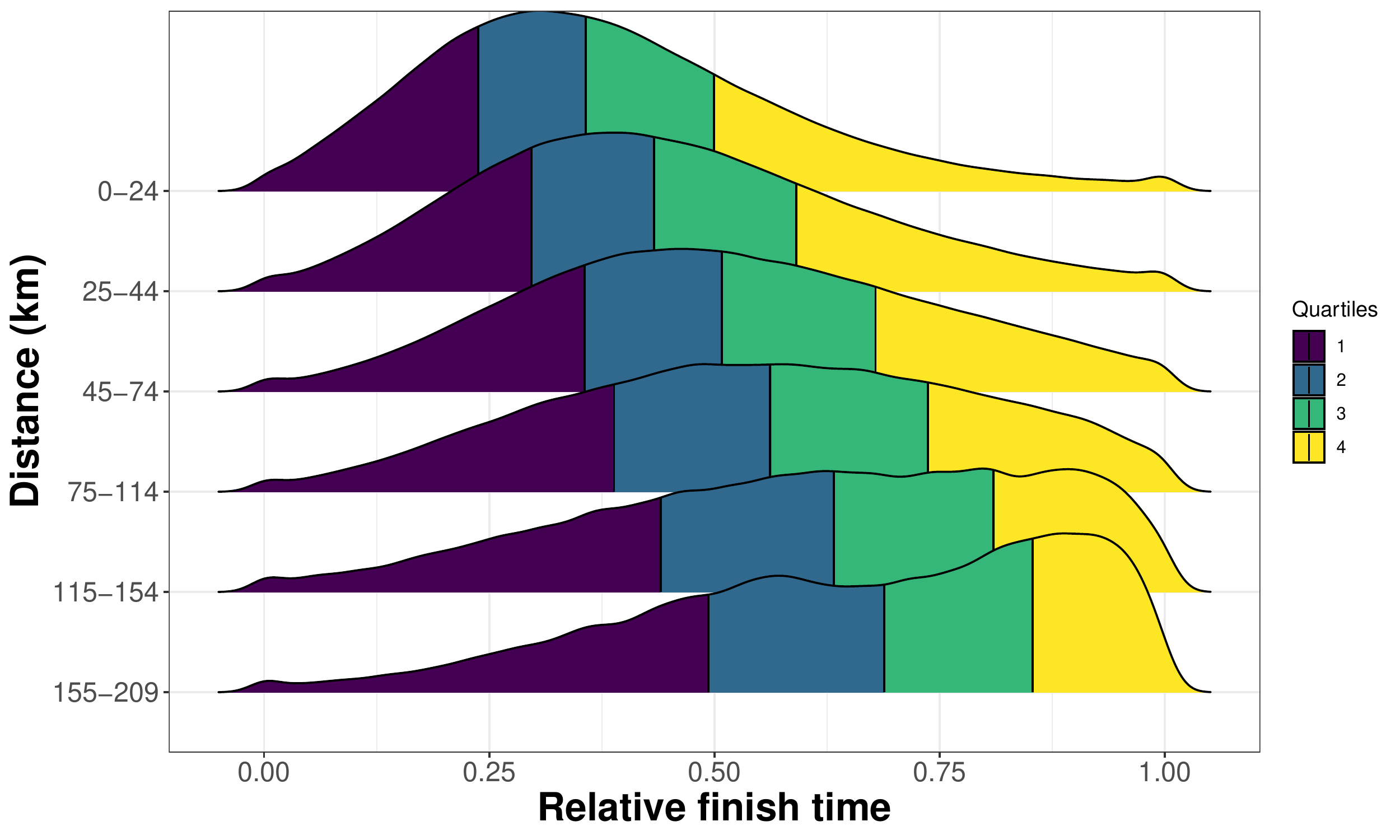}
  \caption{}
  \label{fig:arrival_time_dist}
\end{subfigure} 
\caption{ Data is taken from the ITRA data set. (a) Percentage of women in races by distance category for the years 2010-2018. The percentage is computed as the ratio of the number of women and the total number of participants. Error bars correspond to 95\% confidence intervals for the mean. (b) Distribution of relative finish times as a function of the distance for all races with more than 100 participants. Relative finish time corresponds to the ratio of the difference between finish time and best time in the corresponding race and the difference between worst and best time in the same race. Quartiles are colored according to the legend.}
\end{figure}

We now present two results at the runner-level data. First, female participation appears to be decreasing with distance. Figure~\ref{fig:women_by_distance} shows the proportion of female runners, among all runners, in races belonging to the indicated binned distances. For example, 33\% ($[33\%-34\%]$) of all runners in trail races with distance in the range 0-24 km is female, but this proportion drops to only 22\% ($[21\%-22\%]$) for the distance bin 25-44 km. The fraction of women slightly decreases also for longer distances. Indeed, in the ITRA data set 21\% of all runners are women, and 28\% of them is in races belonging to the 0-24 km, compared to only 15\% of all men running these races. The pattern persists also with respect to elevation and elevation gain per distance. Overall, this suggests that the proportion of female participants is higher in shorter and more flat races. Women's participation also appears to depend on the country of origin: among the five countries listed in Figures~\ref{fig:races_by_month} and~\ref{fig:dist_elev_by_country}, Spain and the US have the lowest and highest number of women in the data set respectively; 13\% ($[13\%-14\%]$) of the participants from Spain are women compared to 34\% ($[34\%-35\%]$). \\
Second, the performance gap between elite and average runners appears to be increasing with distance. Figure~\ref{fig:arrival_time_dist} shows the distribution of relative finish times, which correspond to the ratio of the difference between individual finish time and best time in the corresponding race and the difference between worst and best time in the same race, across distance categories. As distance increases, the shift of the mass towards the right means that the time performance of most runners in races becomes closer the performance of the worst runner in the race. 
For example, if we consider races in the categories 0-24 km and 75-114 km, half of the runners arrive in approximately 35\%  and 56\% of the time gap between the best and worst participants respectively. 
This result may be due to either (1) the fact that most runners slow their pace with distance, while elites do not; or (2) the large fraction of dropouts due to the presence of time barriers in races with longer distances; or (3) selection bias because runners in longer races are more trained, e.g. ITRA points are required.
The literature supporting the first effect in other settings has been discussed in Section~\ref{sec:related}. While we can not check the validity of the three hypotheses in this data set because we do not observe dropouts and partial times, we investigate them in our case study of UTMB in Section~\ref{sec:utmb}. For instance, we observe that also the distribution of arrival times for UTMB is skewed to the left, similarly to the pattern highlighted in Figure~\ref{fig:arrival_time_dist} in the last category.


\end{document}